\documentclass[12pt]{article} 
\usepackage{amstex,amsfonts} \hsize=2in \tolerance=10000
\voffset -0.5in \usepackage{graphicx}
\begin{document}
\title{Long-time-tail Effects on Lyapunov Exponents of a Random,
Two-dimensional Field-driven Lorentz Gas} \author{D. Panja$^{1}$,
J. R. Dorfman$^{1}$, and Henk van Beijeren$^{2}$ \\} \maketitle 
\begin{center}
{\em $^{1}$Institute for Physical Science and Technology and
Department  of Physics,\\ University of Maryland, College Park, MD -
20742, USA \\ $^{2}$Institute for Theoretical Physics, University of
Utrecht \\ Postbus 80006, Utrecht 3508 TA, The Netherlands}
\end{center}
\maketitle

\begin{abstract}
\noindent We study the Lyapunov exponents for a moving, charged
particle in a two-dimensional Lorentz gas with randomly placed,
non-overlapping hard disk scatterers placed in a  thermostatted
electric field, $\vec E $. The low density values of the Lyapunov
exponents have been calculated  with the use of an extended
Lorentz-Boltzmann equation. In this paper we develop a method to
extend these results  to higher density, using the BBGKY hierarchy
equations and extending  them to include the additional variables
needed for calculation of  Lyapunov exponents. We then consider the
effects of correlated collision sequences, due to the so-called ring
events, on the Lyapunov exponents. For small values of the applied
electric field, the ring terms lead to non-analytic, field 
dependent, contributions to both the positive and negative Lyapunov 
exponents which are of the form 
${\tilde{\varepsilon}}^{2}\,\ln\tilde{\varepsilon}$, where 
$\tilde{\varepsilon}$ is a dimensionless parameter  proportional to 
the strength of the applied field. We show that these non-analytic 
terms can be understood as resulting from the change in the collision 
frequency from its equilibrium value, due to the presence of the 
thermostatted field, and that the collision frequency also contains 
such non-analytic terms.  
 
\vspace{5mm} 
\end{abstract} 
 
\vspace{1.5cm} 
\section{Introduction} 
 
\noindent 
The Lorentz gas has proved to be a useful model for studying the 
relations between dynamical systems theory and non-equilibrium 
properties of many body systems. This model consists of a set of 
scatterers that are fixed in space together with moving particles that 
collide  with the scatterers. Here we consider the version of the 
model in two dimensions where the scatterers are fixed hard disks, 
placed at  random in the plane without overlapping. Each of the moving 
particles is a point particle with a mass and a charge, and is 
subjected to an external, uniform electric field as well as a Gaussian 
thermostat which is designed to keep the kinetic energy of the moving 
particle  at a constant value. The particles make elastic, specular 
collisions  with the scatterers, but do not interact with each 
other. The interest in the Lorentz gas model stems from the fact that 
its chaotic properties can be analyzed in some detail, at least if the 
scatterers form a sufficiently dilute, quenched gas, so that the 
average distance between scatterers is large compared to their 
radii. The interest in  a thermostatted electric field arises from the 
fact that at small fields a transport coefficient, the electrical 
conductivity of the particles, is proportional to the sum of the 
Lyapunov exponents describing the chaotic motion of the moving 
particle \cite{EM_ap_book}. The Lyapunov exponents are to be 
calculated for  the case where the charged particle is described by a 
non-equilibrium steady state phase-space distribution function which 
is reached from some typical initial distribution function after a 
sufficiently long period of time. In this state, the distribution 
function for an ensemble of moving particles (all interacting with the 
scatterers and the field, but not with each other) is independent of 
time and its average over the distribution of scatterers is spatially 
homogeneous. It is known from computer simulations 
\cite{EHFML_pra_83,Evans_jcp_83} and  theoretical discussions 
\cite{CELS_cmp_93,Dettmann_preprint} that in the stationary state the 
trajectories of the moving particles in phase space lie on a fractal 
attractor of lower dimension than the dimension of the constant energy 
surface, which is three dimensional for the constant energy Lorentz 
gas in two dimensions. There can be at most two non-zero Lyapunov 
exponents for this model since the Lyapunov exponent in the direction 
of the phase-space trajectory is zero.   Also, the relation between 
the Lyapunov exponents and the electrical conductivity  requires that 
the sum of the non-zero exponents should be negative due to the  
positivity of electrical conductivity \cite{CELS_cmp_93}. 
 
The case of the dilute, random Lorentz gas has already been studied in 
detail. Van Beijeren and coworkers \cite{vBD_prl_95,vBLD_pre_98} have 
calculated the Lyapunov spectrum for an equilibrium Lorentz gas in two 
and three dimensions using various kinetic theory methods including 
Boltzmann equation techniques. These methods were also applied to the 
dilute, random Lorentz gas in a thermostatted electric field with 
results for the Lyapunov exponents that are in excellent agreement 
with computer simulations \cite{vBDCPD_prl_96,LvBD_prl_97}. Moreover, 
the results for the field-dependent case were in accord with the 
relation between the electrical conductivity and the Lyapunov 
exponents for the moving particle.  
 
The purpose of this paper is to extend the results obtained for the
Lyapunov exponents for dilute Lorentz gases to higher densities. Our
central themes will be: (a) to describe a general method, based upon
the BBGKY hierarchy equations, for accomplishing this task, and (b) to
examine the effects on the Lyapunov exponents of long range in time
correlations between the moving particle and the scatterers produced
by correlated collision sequences where the particle collides with a
given scatterer more than once and the time interval between such
re-collisions is on the order of several mean free times, with an
arbitrary number of intermediate collisions with other
scatterers. These correlated collision sequences are of particular
interest in kinetic and transport theory because they are responsible
for the ``long-time-tail'' effects in the Green-Kubo time correlation
functions, which lead to various divergences in the transport
coefficients for two and three dimensional gases, where all of the
particles move \cite{DvB_berne_book}. In the case of a Lorentz gas in
$d$ dimensions, the Green-Kubo velocity correlation functions decay
with time, $t$, as $t^{-(d/2 +1)}$ \cite{EW_pl_71} and the diffusion
coefficient is finite in both two and three dimensions. Here we
describe the effects of these  type of correlations on the Lyapunov
exponents for the two-dimensional Lorentz gas, in equilibrium, where
we find no effect, and in a thermostatted electric field, where we
find a small, logarithmic dependence of the Lyapunov exponents upon
the applied field. This logarithmic effect  is an indicator for
similar effects to be expected when one calculates Lyapunov exponents
associated with more general transport in two-dimensional gases,
whereas in three dimensional systems one would expect corresponding
non-analytic terms proportional to $\tilde{\epsilon}^{\,5/2}$. In the
case of the Lorentz gas, at least, the logarithmic terms can easily be
associated with the logarithmic terms that appear in the field
dependent collision frequency, and a very simple argument can be used
to establish this relation between logarithmic terms in the Lyapunov
exponents and in the collision frequency. 
 
In Section 2 of this article we describe the general theory of
Lyapunov exponents of a two-dimensional thermostatted electric
field-driven  Lorentz gas and quote the results within the scope of
the Boltzmann equation. In Section 3, we generalize the theory to
incorporate the effect of correlated collision sequences. In Section
4, we outline the calculation of the effects of the correlated
collision sequences on the non-zero Lyapunov exponents using the BBGKY
equations discussed in Section 3 and obtain the non-analytic
field-dependent term in the Lyapunov exponents, originating from the
correlated collision sequences, along with other analytic
field-dependent terms. In Section 5, we present some simple arguments
explaining the field-dependence of the collision frequency and show
that this is the sole origin of the non-analytic, field-dependent
terms in the Lyapunov exponents. Notice that the arguments given in
Section 5 are independent of and much simpler to follow than  the
formalism developed in Sections 3 and 4. We conclude  in Section 6
with a discussion of the results obtained here, and with  a
consideration of open questions. Methods for determination of the
field-dependence of the collision frequency are outlined in the
Appendix.  
 
\section{Lyapunov exponents of field driven Lorentz gases in two  
dimensions} 
 
\subsection{General theory} 
 
\noindent The random Lorentz gas consists of point particles of mass 
$m$ and charge $q$ moving in a random array of fixed scatterers. In 
two dimensions, each scatterer is a hard disk of radius $a$. The disks 
do not overlap with each other and are distributed with number density  
$n$, such that at low density $na^{2}<1$. The point particles are acted  
upon by a uniform, constant electric field $\vec{E}$ in the $\hat{x}$  
direction, but there is no interaction between any two point particles.  
There is also a Gaussian thermostat in the system to keep the speed of  
each particle constant at $v$ by means of a dynamical friction during  
flights between collisions with the scatterers. The collisions between  
a point particle and the scatterers are instantaneous, specular and 
elastic. During a flight, the equations of motion of a point particle 
are  
\begin{eqnarray} 
\dot{\vec{r}}\,=\,{\vec{v}}\,=\,\frac{\vec{p}}{m}, {\hspace{1cm}} 
\dot{\vec{p}}\,=\,m\dot{\vec{v}}\,=\,q\vec{E}\,-\, \alpha \vec{p} 
\label{e1} 
\end{eqnarray} 
\hspace{4.9cm}{\includegraphics[width=2.6in]{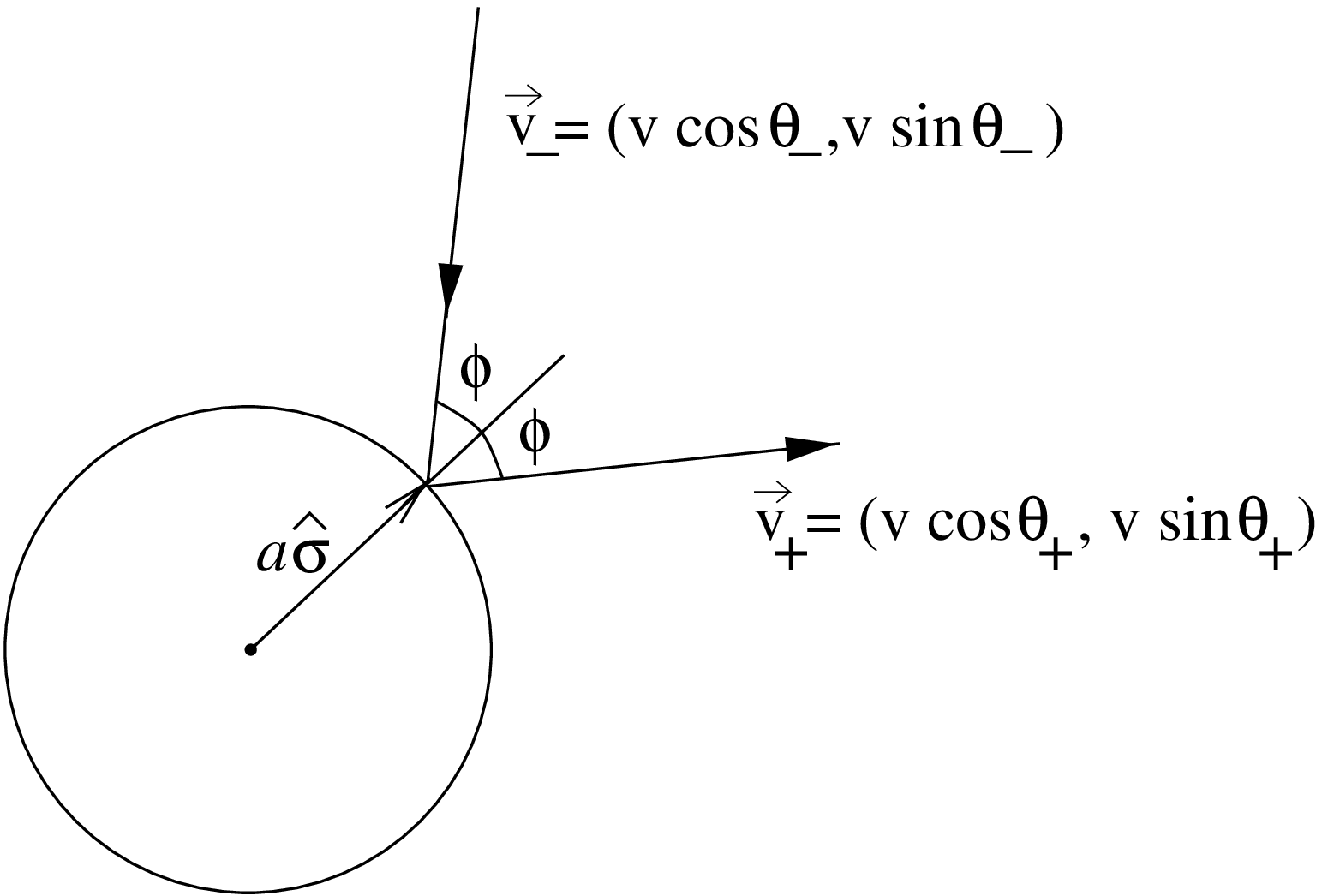}} 
\begin{center} 
Fig. 1 : Collision between a point particle and a scatterer.\\ 
\end{center}  
and at a collision with a scatterer, the post-collisional position and 
velocity, $\vec{r}_{+}$ and $\vec{v}_{+}$, are related to the 
pre-collisional position and velocity, $\vec{r}_{-}$ and 
$\vec{v}_{-}$, by 
\begin{eqnarray} 
\vec{r}_{+}\,=\,\vec{r}_{-}\,, {\hspace{1cm}} 
\vec{v}_{+}\,=\,\vec{v}_{-}\,-\,2\,(\vec{v}_{-}\cdot\hat{\sigma})\,\hat{\sigma}\,, 
\label{e2} 
\end{eqnarray} 
where $\hat{\sigma}$ is the unit vector from the center of the 
scatterer to the point of collision (see Fig. 1). The fact 
that each particle has a constant speed $v$ determines the value of  
$\alpha$ :  
\begin{eqnarray} 
\alpha\,=\,\frac{q\vec{E}\cdot\vec{p}}{p^{2}} 
{\hspace{0.5cm}}\Rightarrow{\hspace{0.5cm}}\dot{\vec{p}}\,=\,q\vec{E}\,-\,\frac{q\vec{E}\cdot\vec{p}}{p^{2}}\,\vec{p}\,. 
\label{e3} 
\end{eqnarray} 
Equivalently, in polar coordinates, the velocity direction with 
respect to the field, defined through
$\hat{v}\cdot\hat{x}\,=\,\cos\theta$, changes between collisions as  
\begin{eqnarray} 
\dot{\theta}\,=\,-\,\varepsilon\sin\theta\,, 
\label{e4} 
\end{eqnarray} 
where $\varepsilon\,=\,\displaystyle{\frac{q |\vec{E}|}{mv}}$ and we 
define the dimensionless electric field parameter 
$\tilde{\varepsilon}\,=\,\displaystyle{\frac{\varepsilon l}{v}}$, 
where $l =(2na)^{-\,1}$ is the mean free path length for the particle 
in the dilute Lorentz gas. To denote the electric field, we will normally 
use $\varepsilon$, though from time to time we will use 
$\tilde{\varepsilon}$, too. 
 
Treating this two-dimensional Lorentz gas as a dynamical system, we 
define the Lyapunov exponents in the usual way: a point particle in 
its phase space $(\vec{r}, \vec{v}) =\vec{{\bf X}}$ starts at time 
$t_{0}$ at a phase space location $\vec{{\bf X}}(t_{0})$. Under time 
evolution, $\vec{{\bf X}}(t)$ follows a trajectory in this phase space 
which we call the ``reference trajectory''. We consider an 
infinitesimally displaced trajectory which starts at the same time 
$t_{0}$, but at $\vec{\bf 
X^{\begin{Sp}\prime\end{Sp}}}(t_{0})\,=\,\vec{{\bf 
X}}(t_{0})\,+\,{\bf\delta\vec{X}}(t_{0})$.  Under time evolution, 
$\vec{{\bf X^{\begin{Sp}\prime\end{Sp}}}}(t)$ follows another 
trajectory, always staying infinitesimally close to the reference 
trajectory. This trajectory we call the ``adjacent 
trajectory''. Typically the two trajectories will separate in time due 
to the convex nature of the collisions. Thus, we can define the 
positive Lyapunov exponent as 
\begin{eqnarray} 
\lambda_{+}\,=\,\lim_{\begin{Sb} T \rightarrow \infty 
\end{Sb}}\,\lim_{\begin{Sb} 
|\delta{\bf\vec{X}}(t_{0})|\rightarrow 0 
\end{Sb}} 
\,\frac{1}{T} \ln 
\frac{|\delta{\bf\vec{X}}(t_{0}\,+\,T)|}{|\delta{\bf\vec{X}}(t_{0})|} 
\,. 
\label{e5} 
\end{eqnarray} 
for a typical trajectory of the system. 
 
We assume that, for small fields, this Lorentz gas system is 
hyperbolic. Since the two-dimensional Lorentz gas can have at most two 
nonzero Lyapunov exponents, we denote the negative Lyapunov exponent 
by $\lambda_{-}$. Without any loss of generality, we can choose to  
measure the separation of the reference and adjacent trajectories  
equivalently in $\vec{r}$-space, thereby reducing the definition of  
the positive Lyapunov exponent to 
\begin{eqnarray} 
\lambda_{+}\,=\,\lim_{\begin{Sb} T \rightarrow \infty 
\end{Sb}}\,\lim_{\begin{Sb} 
|\delta\vec{r}(t_{0})|\rightarrow 0 
\end{Sb}} 
\,\frac{1}{T} \ln 
\frac{|\delta{\vec{r}}(t_{0}\,+\,T)|}{|\delta{\vec{r}}(t_{0})|}\,. 
\label{e6} 
\end{eqnarray} 
In order to calculate the right hand side of Eq. (\ref{e6}), we 
introduce another dynamical quantity, the radius of curvature $\rho$, 
characterizing the spatial separation of the two trajectories (see 
Fig. 2) : 
 
\begin{center}  
\hspace{2cm}{\includegraphics[width=3.67in]{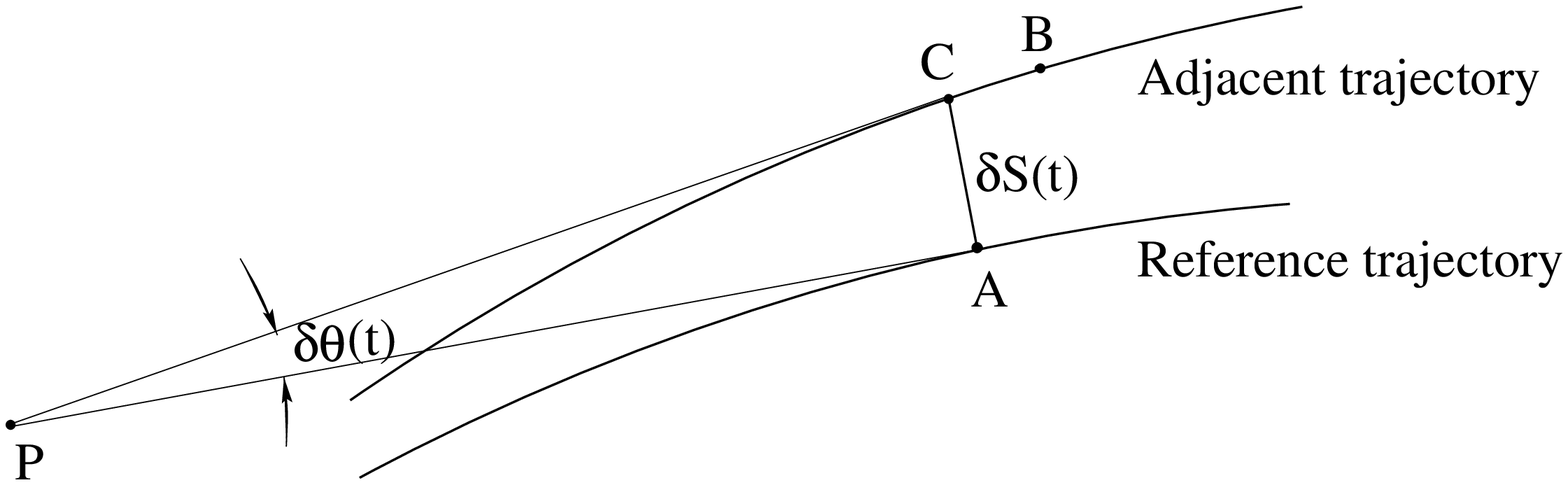}}\\ Fig. 2 : 
$\rho(t)\,=\,\displaystyle{\frac{\delta S(t)}{\delta\theta(t)}}\,=\,|AP|\,$. 
\end{center} 
In Fig. 2, a particle on the reference trajectory would be at point A 
at time $t$. At the same time, a particle on the adjacent trajectory 
would be at B. A local perpendicular on the reference trajectory at A 
intersects the adjacent trajectory at C. The backward extensions of 
instantaneous velocity directions on the reference and adjacent 
trajectories at A and C, respectively, intersect each other at point 
P. We denote the length of the line segment AC by $\delta S(t)$ and 
$\angle APC$ by $\delta\theta(t)$. The radius of curvature associated 
with the particle on the reference trajectory at time $t$ is then 
given by  
\begin{eqnarray} 
\rho(t)\,=\,\frac{\delta S(t)}{\delta \theta (t)}\,=\,|AP|\,. 
\label{e7} 
\end{eqnarray} 
Having defined $\rho(t)$, one can make a simple geometric argument to 
show that  
\begin{equation} 
{\delta\dot{S}}(t)\,=\, v\,{\delta \theta (t)}\,=\,\frac{v\,\delta S(t)}{\rho(t)}, 
\label{e7a} 
\end{equation} 
 so as to obtain a version of Sinai's formula \cite{Sinai_rms_70}, 
\begin{eqnarray} 
\lambda\,=\,\lim_{T\rightarrow\infty}\,\frac{v}{T}\int_{\begin{Sb}t_{0}\end{Sb}}^{\begin{Sp}t_{0}\,+\,T\end{Sp}}\,\frac{dt}{\rho(t)}. 
\label{e8} 
\end{eqnarray} 
 
During a flight, the equation of motion for $\rho$ is given by 
\cite{vBDCPD_prl_96} 
\begin{eqnarray} 
\dot{\rho}\,=\,v\,+\,\rho\varepsilon\cos\theta\,+\,\frac{\rho^{2}\varepsilon^{2}\sin^{2}\theta}{v}\,. 
\label{e9} 
\end{eqnarray} 
At a collision with a scatterer, the post-collisional velocity angle 
$\theta_{+}$ and radius of curvature $\rho_{+}$ are related to the 
pre-collisional velocity angle $\theta_{-}$ and radius of curvature 
$\rho_{-}$ by \cite{Kovacs_priv_disc,Kovacs_preprint} :  
\begin{eqnarray} 
\theta_{+}\,=\,\theta_{-}-\pi+2\phi\,,\hspace{1cm} 
\frac{1}{\rho_{+}}\,=\,\frac{1}{\rho_{-}}\,+\,\frac{2}{a\cos\phi}\,+\,\frac{\varepsilon}{v}\,\tan\phi\,(\sin\theta_{-}\,+\,\sin\theta_{+})\,, 
\label{e10} 
\end{eqnarray} 
where $\phi$ is the collision angle, i.e, 
$\cos\phi\,=\,|\hat{v}_{-}\cdot\hat{\sigma}|\,=\,|\hat{v}_{+}\cdot\hat{\sigma}|$ 
(see Fig. 1).  
 
Now we assume that, for sufficiently weak electric field, the 
field-driven Lorentz gas in two dimensions is ergodic, and that we can 
replace the long time average in Eq. (\ref{e6}) by a non-equilibrium 
steady state (NESS) average, including an average over all allowed 
configurations of scatterers,  to obtain 
\begin{eqnarray} 
\lambda = {\bigg <}\frac{v}{\rho}{\bigg >}_{\mbox{\scriptsize NESS}}\,. 
\label{e11} 
\end{eqnarray} 
The electric field is considered weak if the work done by the electric 
field on the point particle over a flight of  one mean free path is 
much smaller than the particle's kinetic energy, i.e, 
$\displaystyle{\frac{q|\vec{E}|\,l}{mv^{2}}}\,=\,\displaystyle{\frac{\varepsilon l}{v}}\,=\,\tilde{\varepsilon}\,<<\,1$. 
 
We note for future reference, that the sum of the two nonzero Lyapunov 
exponents is related to the average of the friction coefficient 
$\alpha$, by \cite{EM_ap_book,Hoover_elsevier_book,Panja_preprint} 
\begin{equation} 
\lambda_{+}\,+\,\lambda_{-}\,=\,-\,\big<\alpha\big>_{\mbox{\tiny NESS}}\,=\,-\,\bigg<\frac{q\vec{E}\cdot\vec{v}}{mv^{2}}\bigg>_{\mbox{\scriptsize NESS}}\,=\,-\,\frac{\vec{J}\cdot\vec{E}}{mv^{2}}\,=\,-\,\frac{\sigma E^{2}}{mv^{2}}. 
\label{e12} 
\end{equation} 
Here the electric current $\vec{J}\,=\,\langle 
q\vec{v}\rangle_{\mbox{\tiny NESS}}$ is, for small fields,  assumed to 
satisfy Ohm's law, $\vec{J}=\sigma\vec{E}$, and $\sigma$ is the 
electrical conductivity.

\subsection{Results obtained using the Lorentz-Boltzmann equation} 
 
To the lowest order in density, one can assume that the collisions 
suffered by the point particle are uncorrelated, and use an extended 
Lorentz-Boltzmann equation (ELBE) for the distribution function of the 
moving particle, $f_{1}(\vec{r},\vec{v},\rho,t)$ in $(\vec{r}, 
\vec{v}, \rho)$-space \cite{vBDCPD_prl_96} needed for the evaluation 
of the  averages appearing in Eqs. (\ref{e11}) and (\ref{e12}). To 
calculate the positive Lyapunov exponent, one needs to consider the 
forward-time ELBE while to calculate the negative Lyapunov exponent 
one needs the time reversed ELBE. To the leading order in density, the 
Lyapunov exponents are then given by \cite{vBDCPD_prl_96} : 
\begin{eqnarray} 
\lambda^{\mbox{\tiny 
(B)}}_{+}\,=\,\lambda_{0}\,-\,\frac{11}{48}\frac{l}{v}\varepsilon^{2}\,+\,O(\varepsilon^{4})\hspace{1cm}{\mbox{and}}\hspace{1cm}\lambda^{\mbox{\tiny(B)}}_{-}\,=\,-\,\lambda_{0}\,-\,\frac{7}{48}\frac{l}{v}\varepsilon^{2}\,+\,O(\varepsilon^{4})\,. 
\label{e13} 
\end{eqnarray} 
The superscript, B, indicates that these are results obtained from the 
Lorentz-Boltzmann equation. Here $\lambda_{0}$ is the positive 
Lyapunov exponent for a field-free Lorentz gas (see for example 
\cite{D_cup_book}) given by 
\begin{equation} 
\lambda_0 = 2nav\,[\,1 - {\cal{C}} -\ln(2na^{2})\,], 
\label{e131} 
\end{equation} 
where ${\cal{C}}$ is Euler's constant, ${\cal C}=0.5772...$. From  
Eqs. (\ref{e12}) and (\ref{e13}), using Einstein's relation between  
diffusion constant and conductivity, one gets the  correct diffusion  
coefficient within the Boltzmann regime, $D^{\mbox{\tiny (B)}}\,=\,\displaystyle{\frac{3}{8}\,lv}$.  
 
To derive the results in Eq. (\ref{e13}), one uses Eq. (\ref{e10}) 
with $\varepsilon=0$. The $\varepsilon$-dependent term in Eq. 
(\ref{e10}) can be explicitly shown to be of higher order in the 
density than  the terms present in Eq. (\ref{e13}) 
\cite{PD_unpublished}. In the following sections we will investigate 
the effect of sequences of correlated collisions  between the point  
particle and the scatterers on the Lyapunov exponents.  However, 
the $\varepsilon$-dependent term in Eq. (\ref{e10}) will again be 
neglected since we will present the effect of these correlated 
collision sequences in  leading order  in the density of scatterers 
only. Thus, instead of Eq. (\ref{e10}),  we will use 
\begin{eqnarray} 
\frac{1}{\rho_{+}}\,=\,\frac{1}{\rho_{-}}\,+\,\frac{2}{a\cos\phi}\,. 
\label{e14} 
\end{eqnarray}

\section{The extension of the ELBE to higher density} 
 
\subsection{Binary collision operators in $(\vec{r}, \vec{v}, 
\rho)$-space and\\ the BBGKY hierarchy equations}  
 
The Boltzmann theory  for the Lyapunov exponents  assumes that the 
scatterers form a dilute, but quenched system and that the collisions   
of the point particles with the scatterers are  uncorrelated. To 
incorporate the effects of correlated collisions on the Lyapunov 
exponents, we will use a method based on the BBGKY hierarchy equations,  
familiar from the kinetic theory of moderately dense gases  
\cite{DvB_berne_book}. Since the moving particles do not interact with  
each other, it is sufficient to consider the distribution functions  
for just one of them, together with a number of scatterers. One starts  
from a fundamental equation for an $(N+1)$-body distribution function, 
$f_{N+1}\,=\,f_{N+1}(\vec{r}, \vec{v}, \rho; \vec{R}_{1}, 
\vec{R}_{2},.,.,\vec{R}_{N}; t)$, which is the  probability density 
function in the entire  extended phase space $\Gamma$ spanned by  the 
variables $\vec{r}, \vec{v}, \rho, \vec{R}_{1}, 
\vec{R}_{2},.,.,\vec{R}_{N}$, describing our system of $N$ scatterers 
and one moving particle. We  require that $f_{N+1}$ satisfies the 
normalization condition  
\begin{eqnarray} 
\int 
d\vec{r}\,d\vec{v}\,d\rho\,d\vec{R}_{1}\,d\vec{R}_{2}\,.\,.d\vec{R}_{N}\,\,f_{N+1}(\vec{r}, \vec{v}, \rho; \vec{R}_{1}, \vec{R}_{2},.,.,\vec{R}_{N}; t)\,=\,1\,. 
\label{e15} 
\end{eqnarray} 
This $(N+1)$-body distribution function satisfies a  Liouville-like 
equation determined by the collisions of the moving particles with the 
scatterers and by the motion of the particles in the thermostatted 
electric field, between collisions. Since the  time evolution of 
$\vec{r}$ and $\vec{v}$  in this field is not Hamiltonian, we must use 
the Liouville equation in the form of a conservation law, rather than 
the usual form for Hamiltonian systems, to obtain 
\begin{eqnarray} 
\frac{\partial f_{N\,+\,1}}{\partial 
t}\,+\,\vec{\nabla}_{\vec{r}}\cdot(\dot{\vec{r}}\,f_{N\,+\,1})\,+\,\vec{\nabla}_{\vec{v}}\cdot(\dot{\vec{v}}\,f_{N\,+\,1})\,+\,\frac{\partial}{\partial\rho}(\dot{\rho}f_{N\,+\,1})\,=\,\sum_{i\,=\,1}^{N}\tilde{T}_{-,\,i}\,f_{N\,+\,1}\,. 
\label{e22} 
\end{eqnarray} 
Here the operators $\tilde{T}_{-,\,i}$ are binary collision operators 
which describe the effects on the distribution function due to an 
instantaneous, elastic collision between the moving particle and the 
scatterer labeled by the index $i$. The explicit form of the binary 
collision operators may be easily obtained by a slight modification of 
the methods used by Ernst {\it et al.} \cite{DE_jsp_89}, in order to 
include the radius of curvature as an additional variable. One finds 
that the action of this operator on any  function $f(\vec{r}, \vec{v}, 
\rho; \vec{R}_{1}, \vec{R}_{2},.,.,\vec{R}_{j}; t)$ is 
\begin{eqnarray} 
\tilde{T}_{-,\,i}\,\,f\,=\,a\int_{\vec{v}\cdot\hat{\sigma}_{i}\,>\,0}d\hat{\sigma}_{i}\,|\vec{v}\cdot\hat{\sigma}_{i}|\,\bigg\{\int_{0}^{\infty}d\rho'\,\delta\bigg(\rho\,-\,\frac{\rho'\,a\,\cos\phi_{i}}{a\,\cos\phi_{i}\,+\,2\rho'}\bigg)\,\delta(a\hat{\sigma}_{i}\,-\,(\vec{r}\,-\,\vec{R}_{i}))\times\nonumber\\&& 
{\hspace{-8cm}}\times\,b_{\sigma_{i},\,\rho'}\,-\,\delta(a\hat{\sigma}_{i}\,+\,(\vec{r}\,-\,\vec{R}_{i}))\bigg\}\,f\,, 
\label{e17} 
\end{eqnarray} 
where $\hat{\sigma}_{i}$ is the unit vector from the center of the 
scatterer fixed at $\vec{R}_{i}$ to the point of collision. The action 
of the operator $b_{\sigma_{i},\,\rho'}$ on the function $f(\vec{r}, 
\vec{v}, \rho; \vec{R}_{1}, \vec{R}_{2},.,.,\vec{R}_{j}; t)$ is 
defined by  
\begin{eqnarray} 
b_{\sigma_{i},\,\rho'}\,\,f(\vec{r}, \vec{v}, \rho; \vec{R}_{1}, \vec{R}_{2},.,.,\vec{R}_{j}; t)\,=\,f(\vec{r},\,\vec{v}-2\,(\vec{v}\cdot\hat{\sigma}_{i})\,\hat{\sigma}_{i},\,\rho';\vec{R}_{1}, \vec{R}_{2},.,.,\vec{R}_{j}; t)\,; 
\label{e18} 
\end{eqnarray} 
that is, $b_{\sigma_{i}\rho'}$ is a substitution operator that 
replaces  $\rho$ by $\rho'$ and the velocity $\vec{v}$ by its 
restituting  value, i.e, the  value it  should have before collision 
so as to lead to the value $\vec{v}$  after collision.  It is often 
useful to express the binary collision operators as as sum of two 
terms such that 
\begin{eqnarray} 
\tilde{T}_{-,\,i}\,=\,\tilde{T}^{(+)}_{-,\,i}\,-\,\tilde{T}^{(-)}_{-,\,i}\,, 
\label{e21} 
\end{eqnarray} 
where 
\begin{eqnarray} 
\tilde{T}^{(+)}_{-,\,i}\,=\,a\,\int_{\vec{v}\cdot\hat{\sigma}_{i}\,>\,0}d\hat{\sigma}_{i}\,|\vec{v}\cdot\hat{\sigma}_{i}|\,\int_{0}^{\infty}d\rho'\,\delta\bigg(\rho\,-\,\frac{\rho'\,a\,\cos\phi_{i}}{a\,\cos\phi_{i}\,+\,2\rho'}\bigg)\,\delta(a\hat{\sigma}_{i}\,-\,(\vec{r}\,-\,\vec{R}_{i}))\,b_{\sigma_{i},\,\rho'}\, 
\label{e19} 
\end{eqnarray} 
and 
\begin{eqnarray} 
\tilde{T}^{(-)}_{-,\,i}\,=\,a\,\int_{\vec{v}\cdot\hat{\sigma}_{i}\,>\,0}d\hat{\sigma}_{i}\,|\vec{v}\cdot\hat{\sigma}_{i}|\,\delta(a\hat{\sigma}_{i}\,+\,(\vec{r}\,-\,\vec{R}_{i}))\,. 
\label{e20} 
\end{eqnarray} 
One sees that $\tilde{T}^{(+)}_{-,\,i}\,f$ and 
$\tilde{T}^{(-)}_{-,\,i}\,f$ respectively describe the rate of 
``gain''  and the rate of ``loss'' of $f$ due to a collision of the 
point  particle with the scatterer fixed at $\vec{R}_{i}$. 
 
The BBGKY hierarchy equations are then obtained from Eq. (\ref{e22}) 
by integrating over scatterer coordinates, as a set of equations for 
the reduced distributions $f_{j}$ for the moving particle and $(j-1)$ 
scatterers, defined by  
\begin{eqnarray} 
f_{j}(\vec{r}, \vec{v}, \rho; \vec{R}_{1},\vec{R}_{2},.,.,\vec{R}_{j-1};t)\,\nonumber\\&&{\hspace{-2.5cm}}=\,\frac{N!}{(N-j+1)!}\int d\vec{R}_{j}..d\vec{R}_{N}\,f_{N\,+\,1}(\vec{r}, \vec{v}, \rho;\vec{R}_{1}, \vec{R}_{2},.,.,\vec{R}_{N}; t)\,. 
\label{e16} 
\end{eqnarray} 
One then easily obtains the BBGKY hierarchy equations 
($1\,\leq\,j\,\leq\,N$)  
\begin{eqnarray} 
\frac{\partial f_{j}}{\partial 
t}\,+\,\vec{\nabla}_{\vec{r}}\cdot(\dot{\vec{r}}\,f_{j})\,+\,\vec{\nabla}_{\vec{v}}\cdot(\dot{\vec{v}}\,f_{j})\,+\,\frac{\partial}{\partial\rho}(\dot{\rho}f_{j})\,-\,\sum^{j\,-\,1}_{k\,=\,1}\,\tilde{T}_{-,\,k}\,f_{j}\,=\,\int d\vec{R}_{j}\,\tilde{T}_{-,\,j}\,f_{j\,+\,1}\,. 
\label{e23} 
\end{eqnarray} 
 
\subsection{Cluster expansions and truncation of the hierarchy  
equations} 
 
The usual procedure for truncating the hierarchy equations in order to 
obtain the Boltzmann equation and its extension to higher densities is 
to make cluster expansions of the distribution functions, $f_2, 
f_3\,.\,.\,.$ in terms of a set of  correlation functions, $g_2, 
g_3\,.\,.\,.$ as follows: 
\begin{eqnarray} 
f_{2}(\vec{r}, \vec{v}, \rho; \vec{R}_{1}; t)\,=\,nf_{1}(\vec{r}, 
\vec{v}, \rho; t)\,+\,g_{2}(\vec{r}, \vec{v}, \rho; \vec{R}_{1}; t), 
\label{e24} 
\end{eqnarray} 
\begin{eqnarray} 
f_{3}(\vec{r}, \vec{v}, \rho; \vec{R}_{1}, \vec{R}_{2}; 
t)\,=\,n^{2}f_{1}(\vec{r}, \vec{v}, \rho; t)\,+\,ng_{2}(\vec{r}, 
\vec{v}, \rho; \vec{R}_{1}; t)\,+\,ng_{2}(\vec{r}, \vec{v}, \rho; 
\vec{R}_{2}; t)\,\nonumber\\&&{\hspace{-3.7cm}}+\,g_{3}(\vec{r}, 
\vec{v}, \rho; \vec{R}_{1}, \vec{R}_{2}; t)\,,  
\label{e25} 
\end{eqnarray} 
and so on.  Hereafter, to save writing, we denote $g_{2}(\vec{r}, 
\vec{v}, \rho; \vec{R}_{1}; t)$ as $g_{2,\,\vec{R}_{1}}$, 
$g_{2}(\vec{r}, \vec{v}, \rho; \vec{R}_{2}; t)$ as 
$g_{2,\,\vec{R}_{2}}$, $f_{3}(\vec{r},  \vec{v}, \rho; \vec{R}_{1}, 
\vec{R}_{2}; t)$ as $f_{3}$ and $g_{3}(\vec{r}, \vec{v}, \rho; 
\vec{R}_{1}, \vec{R}_{2}; t)$ as $g_{3}$. The first terms in each of 
these expansions represent the totally uncorrelated situation, where 
there are independent probabilities of finding the moving particle and 
the scatterers at the designated coordinates. The next terms involving 
the pair correlation functions $g_{2,\,\vec{R}_{i}}$ in Eqs. 
(\ref{e24}) and (\ref{e25}) take into account possible dynamical and 
excluded volume correlations between the point particle and the 
scatterer at $\vec{R}_{i}$.  If one replaces $f_2$ by $nf_1$ in the 
first BBGKY hierarchy equation, Eq. (\ref{e23}) with $j=1$, reduces to 
the ELBE. To find the corrections to the ELBE for higher 
densities, one must keep the $g_2$ term in Eq. (\ref{e24}) and use the 
second hierarchy equation to determine $g_2$. However, in order to 
solve the second equation, we have to say something about $g_3$.  A 
careful examination of the second and higher equations shows that 
$g_3$ contains, of course, the effects of three-body correlations, 
i.e, correlated collisions involving the point particle, a scatterer 
fixed at $\vec{R}_{1}$ and another scatterer fixed at $\vec{R}_{2}$, 
as  well as excluded volume corrections due to the non-overlapping 
property of the scatterers. Here we will be primarily interested in 
the effects  of the so called ``ring'' collisions on the Lyapunov 
exponents. These collision sequences are composed of one collision of 
the moving particle with a given scatterer, followed by an arbitrary 
number of collisions with a succession of different scatterers, and 
completed  by a final re-collision of the moving particle with the 
first scatterer in the sequence, as illustrated in Fig 3.

\begin{center} 
\unitlength=0.1mm 
\begin{picture}(1650,400)(0,0) 
\thicklines \put(80,300){\circle{200}} \put(134,100){\circle{200}} 
\put(21.72,109.44){\line(1,2){60}} 
\put(21.72,109.44){\vector(1,2){45}} 
\put(83.39,226.77){\line(1,-2){29}} 
\put(83.39,226.77){\vector(1,-2){20}} 
\put(113.36,165.83){\line(0,1){72}} 
\put(113.36,165.83){\vector(0,1){60}} 
\put(113.36,237.83){\line(2,-1){110}} 
\put(113.36,237.83){\vector(2,-1){80}} \put(71,287){\Large{1}} 
\put(125,87){\Large{2}} \put(240,200){\Large {$+$}} 
\put(390,300){\circle{200}} \put(390,230){\line(-1,-1){80}} 
\put(355,195){\vector(1,1){20}} \put(390,230){\line(1,-1){49}} 
\put(390,230){\vector(1,-1){35}} \put(439,110){\circle{140}} 
\put(439,180){\line(1,1){60}} \put(439,180){\vector(1,1){40}} 
\put(571,240){\circle{200}} \put(501,240){\line(-1,1){43}} 
\put(500,240){\vector(-1,1){35}} \put(459,283){\line(2,1){140}} 
\put(459,283){\vector(2,1){60}} \put(381,287){\Large{1}} 
\put(430,97){\Large{2}} \put(562,227){\Large{3}} \put(670,200){\Large 
{$+$}} \put(825,238){\circle{200}} \put(816,225){\Large 1} 
\put(825,166){\line(-2,-1){120}} \put(735,121){\vector(2,1){70}} 
\put(825,166){\line(2,-1){50}} \put(825,166){\vector(2,-1){40}} 
\put(877,70){\circle{200}} \put(867,57){\Large 2} 
\put(877,142){\line(2,1){81}} \put(877,142){\vector(2,1){55}} 
\put(1017,143){\circle{200}} \put(1007,130){\Large 3} 
\put(958,181){\line(0,1){110}} \put(958,181){\vector(0,1){70}} 
\put(1010,337){\circle{200}} \put(1000,325){\Large 4} 
\put(960,287){\line(-1,0){83}} \put(960,287){\vector(-1,0){60}} 
\put(876.5,286.5){\line(0,1){120}} \put(876.5,286.5){\vector(0,1){75}} 
\put(1090,200){\Large {+ .... = }} \put(1480,250){\Large 1} 
\end{picture} 
\end{center} 
\vspace{-4cm} \hspace{13.2cm} 
{\includegraphics[width=1.267in]{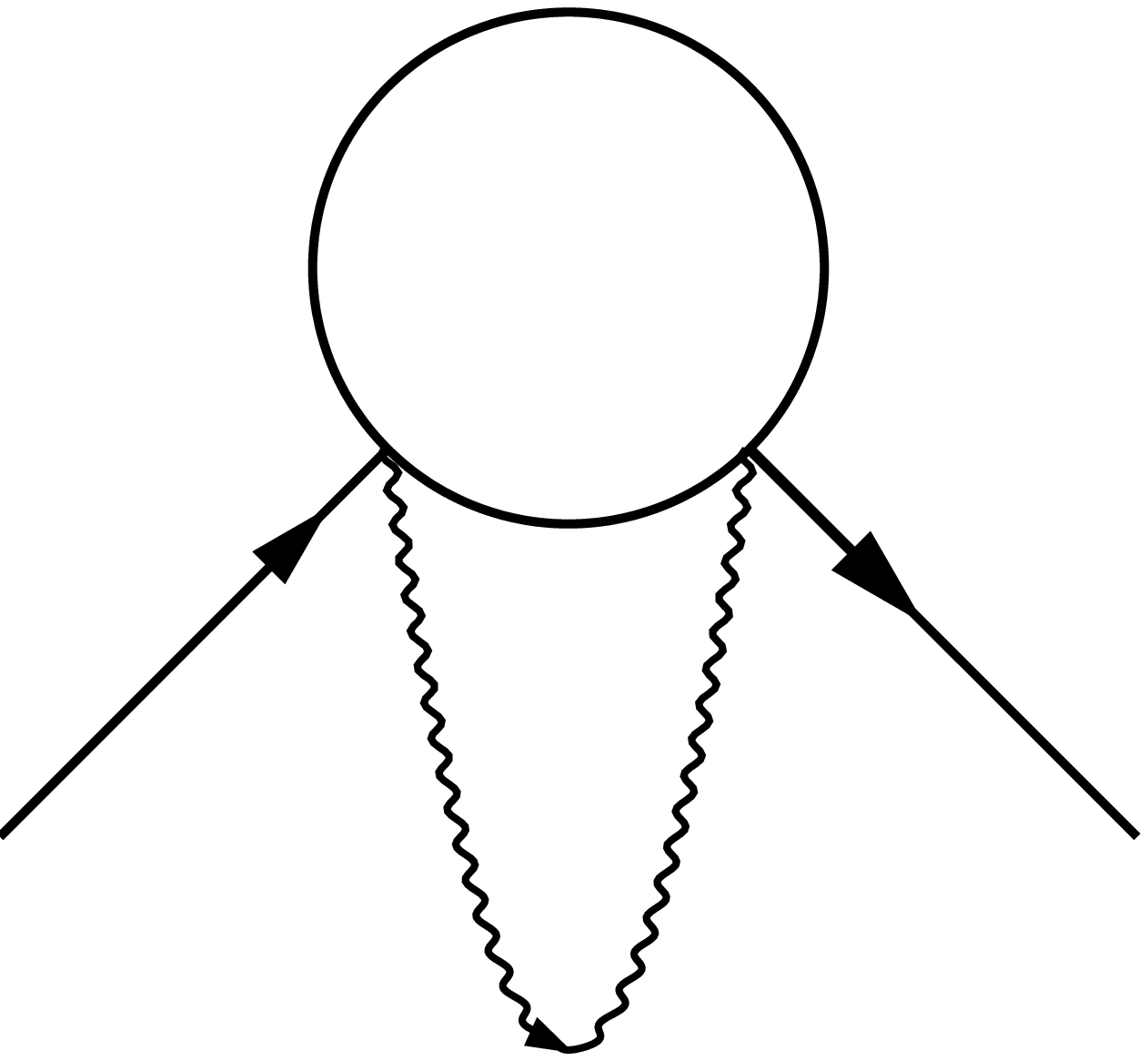}} 
 
\begin{center} 
Fig. 3 : Sequential collisions with scatterers at $\vec{R}_{2},\vec{R}_{3},.,.$ adding up to the ring diagram. 
\end{center} 
 
The ring diagrams, taken individually, are the most divergent terms 
that appear in the expansion of dynamical properties of the Lorentz 
gas as a {\it power} series in the density of scatterers. They lead to 
the logarithmic terms in the density expansion of the diffusion 
coefficient of the moving particle \cite{vLW_physica_67}, and to the 
algebraic  long time tails in the velocity time correlation function 
of the moving particle \cite{EW_pl_71}. While many other dynamical 
events and excluded volume effects contribute to the Lyapunov 
exponents, and  must be included for a full treatment, we concentrate 
here on the effects of these most divergent collision sequences, since 
in other contexts, they are responsible for the most dramatic higher 
density corrections to the Boltzmann equation results.  
 
Thus we drop $g_{3}$ in Eq. (\ref{e25}) and obtain a somewhat 
simplified cluster expansion of $f_{3}$, given by 
\begin{eqnarray} 
f_{3}\,=\,n^{2}f_{1}\,+\,ng_{2,\,\vec{R}_{1}}\,+\,ng_{2,\,\vec{R}_{2}} 
\label{e26} 
\end{eqnarray} 
Using Eqs. (\ref{e24}) and (\ref{e26}) and the first two of the  BBGKY 
hierarchy equations, we obtain a closed set of two equations involving 
two unknowns, $f_{1}$ and $g_{2,\,\vec{R}_{1}}$ given by  
 
\begin{eqnarray} 
\vec{\nabla}_{\vec{v}}\cdot(\dot{\vec{v}}\,f_{1})\,+\,\frac{\partial}{\partial\rho}(\dot{\rho}\,f_{1})\,=\,\int d\vec{R}_{1}\,\tilde{T}_{-,\,1}\,[\,nf_{1}\,+\,g_{2,\,\vec{R}_{1}}]\, 
\label{e33} 
\end{eqnarray} 
and 
\begin{eqnarray} 
\vec{\nabla}_{\vec{r}}\cdot(\dot{\vec{r}}\,g_{2,\,\vec{R}_{1}})\,+\,\vec{\nabla}_{\vec{v}}\cdot(\dot{\vec{v}}\,g_{2,\,\vec{R}_{1}})\,+\,\frac{\partial}{\partial\rho}(\dot{\rho}\,g_{2,\,\vec{R}_{1}})\,-\,n\int d\vec{R}_{2}\,\tilde{T}_{-,\,2}\,g_{2,\,\vec{R}_{1}}\,=\,n\,\tilde{T} 
_{-,\,1}\,f_{1}\,. 
\label{e34} 
\end{eqnarray} 
 
In the derivation of Eq. (\ref{e34}) from the second hierarchy 
equation not only  have we dropped $g_3$ as discussed above, we also 
dropped a term of the form 
$\tilde{T}_{-,\,1}\,g_{2,\,\vec{R_1}}$. This term provides ``repeated 
ring'' corrections to the ring contributions to 
$g_{2,\,\vec{R_1}}$. These  are of the same order as terms neglected 
by dropping $g_{3}$ \cite{ED_physica_72,PD_unpublished}, and should be 
neglected for consistency. We also dropped the time derivatives in the 
equations, so we are now looking for the distribution and correlation 
functions appropriate for the NESS.  
 
In Section 4, we will solve Eqs. (\ref{e33}) and (\ref{e34})  in order 
to calculate the ring contributions to the positive Lyapunov exponent. 
Before doing so, it is useful to write down the usual form of the ring 
equations in $(\vec{r}, \vec{v})$-space, which can be obtained by 
integrating Eqs. (\ref{e33}) and (\ref{e34}) over all values of the 
radius of curvature, $0\leq \rho < \infty$.  We define the usual 
single-particle distribution function by, 
$F_{1}=\int_{\rho>0}d\rho\,f_{1}$ and the pair-correlation function 
$G_{2,\,\vec{R}_{1}}=\int_{\rho>0}d\rho\,g_{2,\,\vec{R}_{1}}$.  By 
imposing the boundary conditions that both $f_{1}$ and 
$g_{2,\,\vec{R}_{1}}$ go to zero as $\rho\rightarrow 0$ and as 
$\rho\rightarrow\infty$, we obtain 
\begin{eqnarray} 
\vec{\nabla}_{\vec{v}}\cdot(\dot{\vec{v}}\,F_{1})\,=\,\int d\vec{R}_{1}\,\overline{T}_{-,\,1}\,[\,nF_{1}\,+\,G_{2,\,\vec{R}_{1}}]\, 
\label{e35} 
\end{eqnarray} 
and  
\begin{eqnarray} 
\vec{\nabla}_{\vec{r}}\cdot(\dot{\vec{r}}\,G_{2,\,\vec{R}_{1}})\,+\,\vec{\nabla}_{\vec{v}}\cdot(\dot{\vec{v}}\,G_{2,\,\vec{R}_{1}})\,-\,n\int d\vec{R}_{2}\,\overline{T}_{-,\,2}\,G_{2,\,\vec{R}_{1}}\,=\,n\,\overline{T}_{-,\,1}\,F_{1}\,. 
\label{e36} 
\end{eqnarray} 
The actions of $\overline{T}_{-,\,1}$ or $\overline{T}_{-,\,2}$ on 
$F_{1}$ and $G_{2,\,\vec{R}_{1}}$ can be obtained by appropriately 
integrating $\tilde{T}_{-,\,1}\,f_{1}$, 
$\tilde{T}_{-,\,1}\,g_{2,\,\vec{R}_{1}}$ or 
$\tilde{T}_{-,\,2}\,g_{2,\,\vec{R}_{1}}$ over $\rho$ from $0$ to 
$\infty$ using the definitions in Eqs. (\ref{e17}) and 
(\ref{e18}). $\overline{T}_{-,\,1}$  and $\overline{T}_{-,\,2}$ are 
the analogs  in ($\vec{r},\,\vec{v}$) space of $\tilde{T}_{-,\,1}$ and 
$\tilde{T}_{-,\,2}$  (see Eqs. (\ref{e17}) and (\ref{e18})), i.e., 
\begin{eqnarray} 
\overline{T}_{-,i}\,=\,a\int_{\vec{v}\cdot\hat{\sigma}_{i}\,>\,0}d\hat{\sigma}_{i}\,|\vec{v}\cdot\hat{\sigma}_{i}|\,\bigg\{\,\delta\,(a\hat{\sigma}_{i}\,-\,(\vec{r}\,-\,\vec{R}_{i}))\,b_{\sigma_{i}}\,-\,\delta(a\hat{\sigma}_{i}\,+\,(\vec{r}\,-\,\vec{R}_{i}))\bigg\}\,. 
\label{e37} 
\end{eqnarray} 
In future applications however, we will drop the $a\hat{\sigma}_{i}$ 
terms from the arguments of both 
$\delta\,(a\hat{\sigma}_{i}\,\pm\,(\vec{r}\,-\,\vec{R}_{i}))$ in 
$\tilde{T}_{-,\,i}$ and $\overline{T}_{-,\,i}$ operators since they 
lead to corrections similar to excluded volume terms, neglected already. 
 
\section{Effects of long range time correlation on $\lambda_{+}$ and $\lambda_{-}$} 
 
We now concentrate on the solution of the BBGKY equations for the 
distribution functions that determine the Lyapunov exponents. The 
solutions of Eqs. (\ref{e33}) and (\ref{e34}) are to be obtained as 
expansions in two small variables, $na^2$ and 
$\tilde{\varepsilon}$. The density expansion will give the corrections 
to the previously obtained Boltzmann regime results from the ELBE, and 
the $\tilde{\varepsilon}$ expansion will provide the field dependence 
of these corrections. We therefore write the density expansions of 
$f_{1}$ and $g_{2}$ (hereafter we drop the subscript $\vec{R}_{1}$ 
from $g_{2,\,\vec{R}_{1}}$) as 
\begin{eqnarray} 
f_{1}\,=\,f_{1}^{\mbox{\tiny(B)}}\,+\,f_{1}^{\mbox{\tiny(R)}}\,+\,.\,.\,. 
\hspace{1cm}{\mbox{and}}\hspace{1cm}g_{2}\,=\,g_{2}^{\mbox{\tiny{(R)}}}\,+\,.\,.\,.\,,  
\label{e38} 
\end{eqnarray} 
where the superscript B indicates the lowest density result for 
$f_{1}$ as given by the ELBE, and the superscript R denotes the ring 
contribution. At the order in density of interest here, the quantities 
indicated explicitly in the above equations satisfy 
\begin{eqnarray} 
\vec{\nabla}_{\vec{v}}\cdot(\dot{\vec{v}}\,f^{\mbox{\tiny{(B)}}}_{1})\,+\,\frac{\partial}{\partial\rho}(\dot{\rho}\,f^{\mbox{\tiny(B)}}_{1})\,=\,n\int d\vec{R}_{1}\,\tilde{T}_{-,\,1}\,f^{\mbox{\tiny (B)}}_{1}, 
\label{e39} 
\end{eqnarray} 
\begin{eqnarray} 
\vec{\nabla}_{\vec{v}}\cdot(\dot{\vec{v}}\,f^{\mbox{\tiny(R)}}_{1})\,+\,\frac{\partial}{\partial\rho}(\dot{\rho}\,f^{\mbox{\tiny{(R)}}}_{1})\,=\,\int d\vec{R}_{1}\,\tilde{T}_{-,\,1}\,[\,nf^{\mbox{\tiny{(R)}}}_{1}\,+\,g^{\mbox{\tiny(R)}}_{2}] 
\label{e40} 
\end{eqnarray} 
and 
\begin{eqnarray} 
\vec{\nabla}_{\vec{r}}\cdot(\dot{\vec{r}}\,g^{\mbox{\tiny{(R)}}}_{2})\,+\,\vec{\nabla}_{\vec{v}}\cdot(\dot{\vec{v}}\,g^{\mbox{\tiny{(R)}}}_{2})\,+\,\frac{\partial}{\partial\rho}(\dot{\rho}\,g^{\mbox{\tiny(R)}}_{2})\,-\,n\int d\vec{R}_{2}\,\tilde{T}_{-,\,2}\,g^{\mbox{\tiny(R)}}_{2}\,=\,n\,\tilde{T}_{-,\,1}\,f^{\mbox{\tiny(B)}}_{1}\,. 
\label{e41} 
\end{eqnarray} 
Our aim here is to solve Eqs. (\ref{e40}) and (\ref{e41}) using the 
results of the ELBE for $f^{\mbox{\tiny (B)}}_1$. We suppose further 
that each of these functions possesses an expansion in powers of 
$\varepsilon$ as 
\begin{eqnarray} 
f^{\mbox{\tiny (B,\,R)}}_{1}&=&f^{\mbox{\tiny(B,\,R)}}_{1,\,0}\,+\,\varepsilon\,f^{\mbox{\tiny(B,\,R)}}_{1,\,1}\,+\,\varepsilon^{2}\,f^{\mbox{\tiny(B,\,R)}}_{1,\,2}\,+\,.\,.\,. 
\label{e42} 
\end{eqnarray} 
and 
\begin{eqnarray} 
g^{\mbox{\tiny (R)}}_{2}&=&g^{\mbox{\tiny(R)}}_{2,\,0}\,+\,\varepsilon\,g^{\mbox{\tiny(R)}}_{2,\,1}\,+\,\varepsilon^2\,g^{\mbox{\tiny(R)}}_{2,\,2}\,+\,.\,.\,.\, 
\label{e43} 
\end{eqnarray} 
The functions $f^{\mbox{\tiny (B)}}_{1,\,i}$ have been previously 
obtained as the  $\varepsilon$ solutions of the ELBE. Since we will be 
dealing with $g_{2}$ only in the context of of the ring term, we drop 
the superscript R from now on. 
 
As mentioned above, we will neglect the term $a\hat{\sigma}$ within 
the arguments of the $\delta$-functions appearing in each of the 
binary collision operators $\tilde{T}_{-}$ and $\overline{T}_{-}$, 
so as to take the moving particle to be located at the same point as 
the center of the appropriate scatterer at collision. The terms 
neglected by this approximation lead to higher density corrections to 
the terms we will obtain below. Secondly, an inspection of the radius 
of curvature delta function in the expression for the ``gain'' part of 
the binary collision operator, Eq. (\ref{e19}), shows that this term 
is only non-vanishing when $\rho \leq \displaystyle{\frac{a}{2}}$, and 
that the dominant contribution to the $\rho'$ integration comes from 
the region $\rho' \sim l$. Naturally, 
$\displaystyle{\frac{\rho'\,a\,\cos\phi_{i}}{a\,\cos\phi_{i}\,+\,2\rho'}\sim\frac{a\,\cos\phi_{i}}{2}(1+O(n))}$ 
in the argument of the delta function. In the Boltzmann level 
approximation this $O(n)$ term may therefore be neglected and it can 
be shown not to contribute to the leading field-dependent ring term 
effects on the Lyapunov exponents. Therefore, we will neglect it in 
what follows. Under this approximation, the gain part of the binary 
collision operator \cite{vBDCPD_prl_96} acts on an arbitrary function 
$h(\vec{r},\vec{v},\rho)$ as  
\begin{eqnarray} 
\tilde{T}^{(+)}_{-,\,i}\,h(\vec{r},\vec{v},\rho)\,\approx\,\delta(\vec{r}\,-\,\vec{R}_i)\,\Theta\big(\frac{a}{2}\,-\,\rho\big)\,I(\rho)\,[\,H(\vec{r},\vec{v}_{+})\,+\,H(\vec{r},\vec{v}_{-})\,]\,\equiv\,\delta(\vec{r}-\vec{R}_{i})\,\Gamma(\rho, H)\,. 
\label{e1new} 
\end{eqnarray} 
Here  
\begin{eqnarray} 
\vec{v}_{\pm}\,=\,\vec{v}\,-\,2\,(\vec{v}\cdot\hat{\sigma}_{i,\,\pm})\,\hat{\sigma}_{i,\,\pm}\,, 
\label{e2new} 
\end{eqnarray} 
and $\hat{\sigma}_{i,\,\pm}$ is defined by the condition that the 
scattering angle 
$\phi=\pm\cos^{-1}\bigg(\displaystyle{\frac{2\rho}{a}}\bigg)$. Also 
\begin{eqnarray} 
H(\vec{r},\vec{v})\,=\,\int_{0}^{\infty} d\rho'\,h(\vec{r}, \vec{v}, 
\rho') 
\label{e3new} 
\end{eqnarray} 
and  
\begin{equation} 
I(\rho)\,=\,\frac{4v\rho}{a\,\sqrt{1 -\big(\frac{2\rho}{a}\big)^{2}}}. 
\label{e4new} 
\end{equation}  
Finally, we express the velocity vector $\vec{v}$ in terms of the 
angle $\theta$ that it makes with the direction of the electric field 
so as to obtain the following set of equations for the terms in the 
$\varepsilon$-expansion of $g_{2}$ 
\begin{eqnarray} 
\bigg[\,\vec{v}\cdot\vec{\nabla}_{\vec{r}}\,+\,v\frac{\partial}{\partial\rho}\,+\,2nav\,\bigg]\,g_{2,\,0}\,=\,-\,2nav\,\delta(\vec{r}\,-\,\vec{R}_{1})\,f^{\mbox{\tiny(B)}}_{1,\,0}\,+\,n\,\delta(\vec{r}-\vec{R}_{1})\,\Gamma(\rho,\,F^{\mbox{\tiny (B)}}_{1,0})\,, 
\label{e44} 
\end{eqnarray} 
\begin{eqnarray} 
\bigg[\,\vec{v}\cdot\vec{\nabla}_{\vec{r}}\,+\,v\frac{\partial}{\partial\rho}\,+\,2nav\,\bigg]\,g_{2,\,1}\,=\,-\,\frac{\partial}{\partial\rho}\bigg(\rho\cos\theta\,g_{2,\,0}\bigg)\,+\,\frac{\partial}{\partial\theta}\bigg(\sin\theta\,g_{2,\,0}\bigg)\,\nonumber \\ 
&&\hspace{-7.5cm}-\,2nav\,\delta(\vec{r}\,-\,\vec{R}_{1})\,f^{\mbox{\tiny(B)}}_{1,\,1}\,+\,n\,\delta(\vec{r}-\vec{R}_{1})\,\Gamma(\rho,F^{\mbox{\tiny(B)}}_{1,1})\,, 
\label{e45} 
\end{eqnarray} 
and 
\begin{eqnarray} 
\bigg[\,\vec{v}\cdot\vec{\nabla}_{\vec{r}}\,+\,v\frac{\partial}{\partial\rho}\,+\,2nav\,\bigg]\,g_{2,\,2}\,=\,-\,\frac{\partial}{\partial\rho}\bigg(\rho\cos\theta\,g_{2,\,1}\,+\,\frac{\rho^2\sin^{2}\theta}{v}\,g_{2,\,0}\bigg)\,+\,\frac{\partial}{\partial\theta}\bigg(\sin\theta g_{2,\,1}\bigg)\,\nonumber 
\\&&\hspace{-9.5cm}-\,2nav\,\delta(\vec{r}\,-\,\vec{R}_{1})\,f^{\mbox{\tiny(B)}}_{1,\,2}\,+\,n\,\delta(\vec{r}-\vec{R}_{1})\,\Gamma(\rho,F^{\mbox{\tiny(B)}}_{1,2})\,. 
\label{e46} 
\end{eqnarray} 
We notice that the equations thus obtained are linear inhomogeneous 
differential equations of the form $L\,g_{2,\,j}=b_{j}$ ($j=0,1,2$), 
where 
$L\,=\,\big[\,\vec{v}\cdot\vec{\nabla}_{\vec{r}}\,+\,v\displaystyle{\frac{\partial}{\partial\rho}}\,+\,2nav\,\big]$ 
is a linear differential operator and $b_{j}$ for $j=0,1,2$ are the 
inhomogeneous terms on the r.h.s. of Eqs. (\ref{e44}), (\ref{e45}) 
and(\ref{e46}) respectively. 
 
We will need to solve Eqs. (\ref{e44}-\ref{e46}), in conjunction with 
the equations for $G_{2,\,0}$, $G_{2,\,1}$ and $G_{2,\,2}$ obtained by 
directly integrating Eqs. (\ref{e44}-\ref{e46}) over $\rho$ from 0 to 
$\infty$. The equations for the corresponding $G_{2,\,i}$ are then  
\begin{eqnarray} 
\bigg[\vec{v}\cdot\vec{\nabla}_{\vec{r}}\,-\,n\int 
\vec{dR_{2}}\,\overline{T}_{-,\,2}\bigg]\,G_{2,\,0}&=&n\,\overline{T}_{-,\,1}\,F^{\mbox{\tiny (B)}}_{1,\,0}\,, 
\label{e47} 
\end{eqnarray} 
\begin{eqnarray} 
\bigg[\vec{v}\cdot\vec{\nabla}_{\vec{r}}\,-\,n\int\vec{dR_{2}}\,\overline{T}_{-,\,2}\bigg]\,G_{2,\,1}&=&\frac{\partial}{\partial\theta}\,\bigg(\sin\theta\,G_{2,\,0}\bigg)\,+\,n\,\overline{T}_{-,\,1}\,F^{\mbox{\tiny(B)}}_{1,\,1} 
\label{e48} 
\end{eqnarray} 
and 
\begin{eqnarray} 
\bigg[\vec{v}\cdot\vec{\nabla}_{\vec{r}}\,-\,n\int\vec{dR_{2}}\,\overline{T}_{-,\,2}\bigg]\,G_{2,\,2}&=&\frac{\partial}{\partial\theta}\,\bigg(\sin\theta\,G_{2,\,1}\bigg)\,+\,n\,\overline{T}_{-,\,1}\,F^{\mbox{\tiny (B)}}_{1,\,2}\,. 
\label{e49} 
\end{eqnarray} 
The equations for $G_{2,\,0}$, $G_{2,\,1}$ and $G_{2,\,2}$ are also 
linear differential  equations of the form 
$L^{\begin{Sp}\prime\end{Sp}}\,G_{2,\,j}=B_{j}$, where 
$L^{\begin{Sp}\prime\end{Sp}}\,=\,\big[\vec{v}\cdot\vec{\nabla}_{\vec{r}}\,-\,n\int\vec{dR_{2}}\,\overline{T}_{-,\,2}\big]$ and $B_{j}$ for $j\,=\,0, 1, 
2$ are the inhomogeneous terms on the r.h.s. of Eqs. (\ref{e47}), 
(\ref{e48}) and (\ref{e49}) respectively. 
 
To solve these equations, we take Fourier transforms of $g_{2}$ and of 
$G_2$ in the variables $\vec{r}$ and in the velocity angle, 
$\theta$. That is, we define 
$\tilde{g}_{2}(\vec{k})\,=\,\displaystyle{\frac{1}{\sqrt{V}}}\,\int_{V}\vec{dr}\,g_{2}\,e^{-\,i\vec{k}\,\cdot\,(\vec{r}\,-\,\vec{R}_{1})}$ 
and calculate $\tilde{g}_{2,\,0}(\vec{k})$, 
$\tilde{g}_{2,\,1}(\vec{k})$ and $\tilde{g}_{2,\,2}(\vec{k})$ in the 
$\vec{k}$-basis, using periodic boundary conditions.  Similarly, we 
define the $m$-th angular mode of $\tilde{g}_{2}(\vec{k})$ as 
$\tilde{g}^{(m)}_{2}(\vec{k})\,=\,\displaystyle{\frac{1}{\sqrt{2\pi}}}\int d\theta\,e^{-\,im\theta}\,\tilde{g}_{2}(\vec{k})$.  We also 
define $\tilde{G}^{(m)}_{2}(\vec{k})$ in an analogous way. Thus, 
corresponding to Eqs. (\ref{e44}-\ref{e46}) and Eqs. 
(\ref{e47}-\ref{e49}), we have two sets of three equations to be 
solved, one involving $\tilde{g}_{2,\,0}(\vec{k})$, 
$\tilde{g}_{2,\,1}(\vec{k})$ and $\tilde{g}_{2,\,2}(\vec{k})$, and the 
other involving $\tilde{G}_{2,\,0}(\vec{k})$, 
$\tilde{G}_{2,\,1}(\vec{k})$ and $\tilde{G}_{2,\,2}(\vec{k})$, in 
$(\vec{k}, m)$ basis. In this basis, the operators 
$L_{\vec{k}}\,=\,\big[\,i\vec{k}\cdot\vec{v}\,+\,v\displaystyle{\frac{\partial}{\partial \rho}}\,+\,2nav\big]$ and 
$L^{\begin{Sp}\prime\end{Sp}}_{\vec{k}}\,=\,\big[\,i\vec{k}\cdot\vec{v}\,-\,n\int\vec{dR_{2}}\,\overline{T}_{-,\,2}\big]$ 
are both infinite dimensional matrices in $m$-space and both of them 
have non-zero off-diagonal elements due to the term 
$i\vec{k}\cdot\vec{v}$ generated from the operator 
$\vec{v}\cdot\vec{\nabla}_{\vec{r}}\,$. However, it is easily seen 
that these off-diagonal elements are proportional to 
$\delta_{m,\,m+1}$ and $\delta_{m,\,m-1}$ and they are easily treated.  
 
A further simplification can be made by noticing that the schematic 
forms of the solutions are 
$\tilde{g}_{2,\,j}(\vec{k})=[L_{\vec{k}}]^{-1}\,b_{j}(\vec{k})$ and 
$\tilde{G}_{2,\,j}(\vec{k})=[L^{\begin{Sp}\prime\end{Sp}}_{\vec{k}}]^{-1}\,B_{j}(\vec{k})$ 
and hence  the dominant parts of $\tilde{g}_{2,\,j}(\vec{k})$ and 
$\tilde{G}_{2,\,j}(\vec{k})$ will come, loosely speaking, from the 
eigenfunctions of $L_{\vec{k}}$ and 
$L^{\begin{Sp}\prime\end{Sp}}_{\vec{k}}$ having the smallest 
eigenvalues. The lowest eigenvalues of 
$L^{\begin{Sp}\prime\end{Sp}}_{\vec{k}}$ are $\propto k^{2}$ due to 
the contributions from  the hydrodynamic modes \cite{EW_pl_71}. Thus, 
to capture the dominant part of the solutions we should solve the 
equations in the range $k = |\vec{k}| << l^{-1}$, the inverse mean 
free path and use perturbation expansions in the small parameter $kl$. 
We will not, in our analysis, follow the mode expansion technique, as 
it is simpler to calculate $G_{2}$ directly. However, one can use mode 
expansions and one finds that the results of both the methods agree. 
 
\subsection{Solution for $G_{2}$} 
 
\noindent To solve for $G_{2}$ first we need to know the solutions of 
the Lorentz-Boltzmann equation for $F^{\mbox{\tiny (B)}}_{1,\,0}$, 
$F^{\mbox{\tiny (B)}}_{1,\,1}$ and $F^{\mbox{\tiny (B)}}_{1,\,2}$. 
These are given by \cite{vBDCPD_prl_96}  
\begin{eqnarray} 
F^{\mbox{\tiny 
(B)}}_{1,\,0}\,=\,\frac{1}{2\pi},\hspace{0.8cm}F^{\mbox{\tiny(B)}}_{1,\,1}\,=\,\frac{3}{16\pi nav}\,\cos\theta\hspace{0.6cm}{\mbox{and}}\hspace{0.6cm}F^{\mbox{\tiny(B)}}_{1,\,2}\,=\,\frac{45}{512\pi (nav)^{2}}\,\cos 2\theta\,. 
\label{e50} 
\end{eqnarray} 
 
We note that in $m$-space, defined above, the $m$-th diagonal element 
of the infinite matrix $L^{'}_{\vec{k}}$ is 
$\displaystyle{\frac{4m^{2}}{(4m^{2}\,-\,1)}}\displaystyle{\frac{v}{l}}$ 
while the off-diagonal elements are $\,ikv\, \delta_{m,m\pm1}$. Thus 
an expansion in $\tilde{k}=kl$ can be easily obtained by considering 
successively larger parts of the matrix $L^{'}_{\vec{k}}$ in the 
index $m$, starting with $3\times 3$, $5\times 5$ matrices and so on, 
chosen in such a way that the element of $L^{'}_{\vec{k}}$ 
corresponding to $m=0$ appears as the center element of these 
matrices. As we want to make our results correct up to 
$O(\tilde{k}^{0})$, we need to  increase the size of these matrices 
till the expressions of $\tilde{G}_{2,\,j}(\vec{k})$ obtained from 
$\tilde{G}_{2,\,j}(\vec{k})=[L^{\begin{Sp}\prime\end{Sp}}_{\vec{k}}]^{-1}\,B_{j}(\vec{k})$ 
(for $j=0, 1, 2$) converges up to $O(\tilde{k}^{0})$. Also, as we want 
to obtain the expression of $\lambda_{+}$ and $\lambda_{-}$ in the 
leading field-dependent order, which is $ \varepsilon^{2}$; we need 
the solutions of all the $m$-modes of $\tilde{G}_{2,\,0}(\vec{k})$, 
$\tilde{G}_{2,\,1}(\vec{k})$ and $\tilde{G}_{2,\,2}(\vec{k})$ that are 
necessary to obtain all the terms of $f^{\mbox{\tiny (R)}}_{1}$ that 
are $\propto\varepsilon^{2}$ and contribute to this leading 
field-dependent order of $\lambda_{+}$ and $\lambda_{-}$. In more 
explicit form, this means that we definitely need the solutions of 
$\tilde{G}^{(m\,=\,0)}_{2,\,0}(\vec{k})$, 
$\tilde{G}^{(m\,=\,0)}_{2,\,1}(\vec{k})$, 
$\tilde{G}^{(m\,=\,\pm1)}_{2,\,1}(\vec{k})$, 
$\tilde{G}^{(m\,=\,\pm2)}_{2,\,1}(\vec{k})$ and 
$\tilde{G}^{(m\,=\,0)}_{2,\,2}(\vec{k})$ up to 
$O(\tilde{k}^{0})$. However,  once we present these solutions, from 
the structure and properties of them,  it will turn out that we will 
also need the expressions of $\tilde{G}^{(m=\,0)}_{2,\,2j}(\vec{k})$ 
in  the leading order of $\tilde{k}\,$ for $j\,=\,2, 3,\,.\,.$, to 
consistently obtain all the terms, that are $\propto\varepsilon^{2}$.  
 
At the $\varepsilon^{0}$ or equilibrium order, we find 
\begin{eqnarray} 
\tilde{G}^{(m)}_{2,\,0}(\vec{k})\,=\,0 \hspace{1cm}\forall m\,. 
\label{e51} 
\end{eqnarray} 
Proceeding to order $\varepsilon$, we  find that 
$\tilde{G}^{(m)}_{2,\,1}(\vec{k})$'s obey  
\begin{eqnarray} 
\frac{iv}{2}\,\bigg[(k_{x}\,+\,ik_{y})\,\tilde{G}^{(m+1)}_{2,\,1}(\vec{k})\,+\,(k_{x}\,-\,ik_{y})\,\tilde{G}^{(m-1)}_{2,\,1}(\vec{k})\bigg]\,+\,\frac{8navm^{2}}{4m^{2}\,-\,1}\,\tilde{G}^{(m)}_{2,\,1}(\vec{k})\nonumber\\ 
&&\hspace{-4cm}\,=\,-\,\frac{1}{2\,\sqrt{2\pi V}}\,\big(\delta_{m,\,1}\,+\,\delta_{m,\,-1}\big)\,, 
\label{e52} 
\end{eqnarray} 
where it turns out that we need a $5\times 5$ matrix block 
corresponding to $m=-2,\,-1,\,0,\,1,$ and $2$ to get the solutions of 
$\tilde{G}^{(m)}_{2,\,1}(\vec{k})$ up to $O(k^{0})$ that are relevant 
for us, yielding  
\begin{eqnarray} 
\tilde{G}^{(m=\,0)}_{2,\,1}(\vec{k})\,=\,\frac{ik_{x}}{vk^{2}\,\sqrt{2\pi V}}\,,  \nonumber 
\end{eqnarray} 
\begin{eqnarray} 
\tilde{G}^{(m=1)}_{2,\,1}(\vec{k})\,=\,-\,\frac{3ik_{y}(k_{x}\,-\,ik_{y})}{16navk^{2}\,\sqrt{2\pi V}}\,,\hspace{1.5cm}\tilde{G}^{(m=\,-1)}_{2,\,1}(\vec{k})\,=\,\frac{3ik_{y}(k_{x} + ik_{y})}{16navk^{2}\,\sqrt{2\pi V}}  \nonumber 
\end{eqnarray} 
\begin{eqnarray} 
\tilde{G}^{(m=\,2)}_{2,\,1}(\vec{k})\,=\,-\,\frac{45k_{y}(k_{x}\,-\,ik_{y})^{2}}{1024(na)^{2}vk^{2}\,\sqrt{2\pi V}}\hspace{0.7cm}{\mbox{and}}\hspace{0.7cm}\tilde{G}^{(m=\,-2)}_{2,\,1}(\vec{k})\,=\,\frac{45k_{y}(k_{x}\,+\,ik_{y})^{2}}{1024(na)^{2}vk^{2}\,\sqrt{2\pi V}}\,. 
\label{e53} 
\end{eqnarray} 
Notice that we have also calculated 
$\tilde{G}^{(m=\,-\,2)}_{2,\,1}(\vec{k})$ and 
$\tilde{G}^{(m=\,2)}_{2,\,1}(\vec{k})$, even though they are $O(k)$, 
because they affect the $O(k^{0})$ solution for 
$\tilde{G}^{(m=\,0)}_{2,\,2}(\vec{k})$. 
 
For order $\varepsilon^2$, the relevant 
$\tilde{G}^{(m)}_{2,\,2}(\vec{k})$'s are then calculated using Eq. 
(\ref{e53}) and considering a $5\times 5$ matrix block of 
$L^{\begin{Sp}\prime\end{Sp}}_{\vec{k}}$. There we need only the 
solution for $\tilde{G}^{(m=\,0)}_{2,\,2}(\vec{k})$ : 
\begin{eqnarray} 
\tilde{G}^{(m=\,0)}_{2,\,2}(\vec{k})\,=\,\frac{k^{2}_{x}}{v^{2}k^{4}\,\sqrt{2\pi V}}\,+\,\frac{45\,(2k_{x}^{2}\,-\,5k_{y}^{2})}{1024(nav)^{2}k^{2}\,\sqrt{2\pi V}}\,. 
\label{e54} 
\end{eqnarray} 
Examining the properties of the solutions, Eqs. (\ref{e53}-\ref{e54}) 
and observing from Eqs. (\ref{e47}-\ref{e49}) the way the solution of 
$G_{2,\,j}$ affects the solution of $G_{2,\,(j\,+\,1)}$, one sees that 
the leading power of $k$ in the expression of 
$\tilde{G}^{(m=\,0)}_{2,\,2j}(\vec{k})$ for ($j=1, 2, 3....$) is 
$k^{-\,2j}$. However, in the expression of $G_{2}$, 
$\tilde{G}^{(m=\,0)}_{2,\,2j}(\vec{k})$ appears with a factor of 
$\varepsilon^{2j}$. When $G_{2}$ is finally calculated, after a 
summation of the appropriate $\vec{k}$-values\footnote{To see how the 
$\vec{k}$-integration is performed, see the last paragraph of Section 
4.2}, the contribution of the sum of all the effects coming from the 
$O(k^{-\,2j})$ terms of the $\tilde{G}^{(m=\,0)}_{2,\,2j}(\vec{k})$'s 
is seen to be in the same order of density of scatterers as the 
$O(k^{0})$ term on the r.h.s. of Eq. (\ref{e54}). In fact, it also 
turns out that the $O(k^{-\,2j})$ terms of the 
$\tilde{G}^{(m=\,0)}_{2,\,2j}(\vec{k})$'s are the only ones among the 
$\tilde{G}^{(m)}_{2,\,j}(\vec{k})$'s that contribute to 
$f^{\mbox{\tiny (R)}}_{1}$ in the order of $\varepsilon^{2}$. This 
implies that along with the solutions, Eqs. (\ref{e51}), (\ref{e53}) 
and (\ref{e54}), we also need to include the $O(k^{-\,2j})$ term of 
$\tilde{G}^{(m=\,0)}_{2,\,2j}(\vec{k})$ to be consistent. If one just 
considers this $O(k^{-\,2j})$ term in 
$\tilde{G}^{(m=\,0)}_{2,\,2j}(\vec{k})$, then it is easy to see that 
they satisfy a recurrence relation for $j\geq 1$ :  
\begin{eqnarray} 
\tilde{G}^{(m=\,0)}_{2,\,2\,(j\,+\,1)}(\vec{k})&=&-\,\frac{k^{2}_{x}}{v^{2}k^{4}}\,\tilde{G}^{(m=\,0)}_{2,\,2j}(\vec{k})\,, 
\label{e55} 
\end{eqnarray} 
i.e, 
\begin{eqnarray} 
\tilde{G}^{(m=\,0)}_{2,\,2j}(\vec{k})&=&\frac{(-1)^{\,j\,-\,1}}{\sqrt{2\pi V}}\,\bigg(\frac{k^{2}_{x}}{v^{2}k^{4}}\bigg)^{j}\,. 
\label{e56} 
\end{eqnarray}  
The solutions, Eqs. (\ref{e51}), (\ref{e53}), (\ref{e54}) and 
(\ref{e56}), are then used to determine the integration constants that 
arise when we solve the differential Eqs. (\ref{e44}-\ref{e46}).  It 
is important to note that the first term on the right hand side of 
Eq. (\ref{e54}) is inversely proportional to $k^2$. This is the origin 
of the logarithmic terms we find below. 
 
\subsection{Solution for $g_{2}$} 
 
Here we apply the same procedure to solve for the 
$\tilde{g}_{2}(\vec{k})$'s from the equations 
$L_{\vec{k}}\,\tilde{g}_{2}(\vec{k})\,=\,b_{j}(\vec{k})$ for 
$j\,=\,0,\,1,\,2$. This time, the elements of $L_{\vec{k}}$ are 
differential operators in the variable $\rho$ and the corresponding 
constants of integrations are determined using the solutions of 
$\tilde{G}^{(m)}_{2}(\vec{k})$'s while maintaining that 
$\tilde{g}^{(m)}_{2}(\vec{k})$'s go to zero as $\rho\rightarrow 0$ and 
as $\rho\rightarrow\infty$. We also note that for our purpose, 
solutions of the $\tilde{g}_{2}(\vec{k})$'s are only needed for 
$\rho>\displaystyle{\frac{a}{2}}$ as the solution of the 
$\tilde{g}_{2}(\vec{k})$'s for $\rho<\displaystyle{\frac{a}{2}}$ gives 
rise to higher order density corrections than under consideration 
here. These solutions can also be obtained by the mode expansion 
technique discussed above in the paragraph preceding Section 4.1. 
However, as it is fairly straightforward to solve the differential 
Eqs. (\ref{e44}-\ref{e46}) for $\rho>\displaystyle{\frac{a}{2}}$, we 
directly write down the necessary solutions up to $O(k^{0})$.  
 
We obtain, for $\rho>\displaystyle{\frac{a}{2}}$, 
\begin{eqnarray} 
\tilde{g}^{(m)}_{2,\,0}(\vec{k})\,=\,\frac{2na}{\sqrt{2\pi V}}\,(1\,-\,2na\rho)\,e^{-\,2na\rho}\,, \nonumber 
\end{eqnarray} 
\begin{eqnarray} 
\tilde{g}^{(m=\,-1)}_{2,\,1}(\vec{k})\,=\,\frac{1}{v\,\sqrt{2\pi V}}\,\bigg[-\,\frac{k_{x}(k_{x}\,+\,ik_{y})}{8k^{2}}\,+\,\frac{k_{x}(k_{x}\,+\,ik_{y})}{2k^{2}}\,2na\rho\,+\,\frac{2na\rho}{8}\,\nonumber\\  
&& {\hspace{-4cm}}+\,\frac{(2na\rho)^{2}}{2}\,-\,\frac{(2na\rho)^{3}}{4}\,\bigg]\,e^{-\,2na\rho}\,, \nonumber 
\end{eqnarray} 
\begin{eqnarray} 
\tilde{g}^{(m=\,0)}_{2,\,1}(\vec{k})\,=\,\frac{ik_{x}}{vk^{2}\,\sqrt{2\pi V}}\,2na\,e^{-\,2na\rho}\,, \nonumber 
\end{eqnarray} 
\begin{eqnarray} 
\tilde{g}^{(m=\,1)}_{2,\,1}(\vec{k})\,=\,\frac{1}{v\,\sqrt{2\pi V}}\,\bigg[-\,\frac{k_{x}(k_{x}\,-\,ik_{y})}{8k^{2}}\,+\,\frac{k_{x}(k_{x}\,-\,ik_{y})}{2k^{2}}\,2na\rho\,+\,\frac{2na\rho}{8}\,\nonumber\\  
&& {\hspace{-4cm}}+\,\frac{(2na\rho)^{2}}{2}\,-\,\frac{(2na\rho)^{3}}{4}\,\bigg]\,e^{-\,2na\rho}\,, \nonumber 
\end{eqnarray} 
\begin{eqnarray} 
\tilde{g}^{(m=\,0)}_{2,\,2}(\vec{k})\,=\,\frac{k^{2}_{x}}{v^{2}k^{4}\,\sqrt{2\pi V}}\,2na\,e^{-\,2na\rho}\nonumber\\  
&& 
{\hspace{-4.5cm}}+\,\frac{1}{2v^{2}\,\sqrt{2\pi V}}\bigg[\frac{45}{128na}\,-\,\frac{315k_{y}^{2}}{256nak^{2}}\,+\,\frac{k_{x}^{2}}{4k^{2}}\,\rho\,-\,\frac{11}{8}\,2na\rho^{2}\,\nonumber\\ 
&&{\hspace{-1.5cm}}-\,\frac{13}{8}\,\frac{k_{x}^{2}}{k^{2}}\,2na\rho^{2}\,+\,\frac{19}{24}\,(2na)^{2}\rho^{3}\,+\,\frac{k_{x}^{2}}{2k^{2}}\,(2na)^{2}\rho^{3}\,\nonumber\\ 
&& 
+\,\frac{13}{24}\,(2na)^{3}\rho^{4}\,-\,\frac{1}{8}\,(2na)^{4}\rho^{5}\bigg]\,e^{-2na\rho} 
\label{e61} 
\end{eqnarray} 
and for the $O(k^{-\,2j})$ terms in 
$\tilde{G}^{(m=\,0)}_{2,\,2j}(\vec{k})$ we have  
\begin{eqnarray} 
\tilde{g}^{(m=\,0)}_{2,\,2j}(\vec{k})&=&\frac{(-1)^{\,j\,-\,1}}{\sqrt{2\pi 
 V}}\,\bigg(\frac{k^{2}_{x}}{v^{2}k^{4}}\bigg)^{j}\,2na\,e^{-\,2na\rho}\,. 
\label{e62} 
\end{eqnarray} 
We point out that all of the
terms in each of the square brackets, in each of the above three
equations, are of the same order in the density. This can be seen
easily by noting that $\rho$ is typically of order $(2na)^{-1}$, so
that $(2na\rho)$ is typically independent of the density.
The solutions, Eqs. (\ref{e51}), (\ref{e53}), (\ref{e54}), 
(\ref{e56}), (\ref{e61}) and (\ref{e62}) now can be assembled to 
calculate $G_{2}$ and $g_{2}$ in $(\vec{r}, \vec{v})$ and $(\vec{r}, 
\vec{v}, \rho)$ space respectively and  feed the results into the 
r.h.s. of Eqs. (\ref{e35}) and (\ref{e40}) to obtain $f^{\mbox{\tiny(R)}}_{1}$. This involves a summation of different $m$ and 
$\vec{k}$-values.  In the infinite volume limit the $\vec{k}$-sum can 
be converted to an integration over $\vec{k}$. The sum over $m$ is 
straightforward, but we have to remember that the integration over 
$\vec{k}$ has to be carried out in a range $k\leq 
k_{0}\sim l^{-1}$. Secondly, since we have expanded the distribution
functions in powers of $\varepsilon$ and then subsequently in powers
of $k$, the lower limit of $k$ for the $\vec{k}$-integration cannot be
taken to be zero. To determine this lower limit of $k$ for the
$k$-integration, we observe that the expansion in $\varepsilon$ cannot
be carried out for those values of $k$ where
$k<\displaystyle{\frac{\varepsilon}{2v}}$, so that the value
$\displaystyle{\frac{\varepsilon}{2v}}$ forms a natural lower cut-off
for the Fourier transform. Our solutions of $G_{2}$ and $g_{2}$
therefore do not hold for $k<\displaystyle{\frac{\varepsilon}{2v}}$
and to do a satisfactory perturbation theory in the range
$k<\displaystyle{\frac{\varepsilon}{2v}}$, one needs to consider both
the $\varepsilon$ and $\vec{k}$-dependent terms together. After doing
so,  one finds that such a perturbation theory does not affect our
results at the present density order \cite{PD_unpublished}. 
 
Before performing the integration over $\vec{k}$, we notice that in
two dimensions, the numerator of the $\vec{k}$-integral is
proportional to $k\,dk$. This means that any part of the solutions of
$\tilde{g}_{2}(\vec{k})$ or $\tilde{G}_{2}(\vec{k})$ having a  leading
power of $k$ of order 2 or higher in the denominator gives rise to a
singularity at $k\rightarrow 0$ for the $\vec{k}$-integral. First, the
highest leading power of $k$ in the denominators of Eqs. (\ref{e51}),
(\ref{e53}), (\ref{e54}) and (\ref{e61}) is $k^{2}$, occurring in
$\tilde{G}^{(m\,=\,0)}_{2,\,2}(\vec{k})$ and
$\tilde{g}^{(m\,=\,0)}_{2,\,2}(\vec{k})$ respectively. These terms
proportional to $k^{-\,2}$ give rise to a logarithmic electric field
dependence once the $\vec{k}$-integration is performed for
$\displaystyle{\frac{\varepsilon}{2v}}\leq k\leq k_{0}$. The rest of
the terms in these solutions supply only analytic field dependences
that can be expressed as power series in $\varepsilon$. Secondly, even
though the solutions given in Eqs. (\ref{e56}) and (\ref{e62})   have
higher powers of $k$ than $k^{2}$ in the  denominators, they also come
with subsequently higher powers of  $\varepsilon$ in their
numerators. Thus, when the  $\vec{k}$-integration is performed, they
contribute terms proportional  to $\varepsilon^{2}$ or higher, to
$g_{2}$ or $G_{2}$. Consequently,  in addition to analytic field
dependent terms, in our present  approximation we have only one
non-analytic field  dependent term  appearing in $g_{2}$ or $G_{2}$
and that is proportional to
$\tilde{\varepsilon}^{2}\,\ln\,\tilde{\varepsilon}$.  No doubt there
exist further non-analytic terms in higher orders in
$\tilde{\varepsilon}$, but their calculation would require a careful
consideration of  various terms we have  neglected here, such as the
repeated ring contributions. 
 
\subsection{Solution for $f^{\mbox{\tiny (R)}}_{1}$ and the  
calculation of $\lambda^{\mbox{\tiny (R)}}_{+}$} 
 
Once the solutions, Eqs. (\ref{e51}), (\ref{e53}), (\ref{e54}), 
(\ref{e56}), (\ref{e61}) and (\ref{e62}) are inserted in 
Eqs. (\ref{e35}) and (\ref{e41}) and the $\vec{k}$-integration is 
performed in the range 
$\displaystyle{\frac{\varepsilon}{2v}}<k<k_{0}$, we get,  by the 
method described in (\ref{e1new}-\ref{e4new}), the following equations 
to be solved to obtain  
$f^{\mbox{\tiny (R)}}_{1}$ and $F^{\mbox{\tiny(R)}}_{1}$ respectively :  
\begin{eqnarray} 
-\,\varepsilon\,\frac{\partial}{\partial\theta}\,(\sin\theta\,f^{\mbox{\tiny(R)}}_{1})\,+\,\frac{\partial}{\partial\rho}\,\bigg\{\bigg(v\,+\,\rho\varepsilon\cos\theta\,+\,\frac{\rho^{2}\varepsilon^{2}\sin^{2}\theta}{v}\bigg)\,f^{\mbox{\tiny (R)}}_{1}\bigg\}\,+\,2navf^{\mbox{\tiny(R)}}_{1}\nonumber\\ 
&&{\hspace{-14cm}}=\,\Theta\big(\frac{a}{2}\,-\,\rho\big)\,\frac{4v\rho}{a\sqrt{1\,-\,\big(\frac{2\rho}{a}\big)^{2}}}\times\nonumber 
\\&&{\hspace{-13.7cm}}\times\!\int^{\frac{\pi}{2}}_{-\,\frac{\pi}{2}}\!d\phi\,\cos\phi\,b_{\sigma}\bigg[\,\frac{\varepsilon^{2}}{8\pi^{2}v^{2}}\,\bigg\{\ln\frac{2vk_{0}}{\varepsilon}\bigg\}\!-\!\frac{k^{2}_{0}}{8\pi^{2}}\bigg\{\!\frac{3}{16nav}\,\varepsilon\cos\theta\,+\,\frac{135}{2048(nav)^{2}}\varepsilon^{2}\bigg\}\!+\!\frac{A\,\varepsilon^{2}}{16\pi^{3}v^{2}}\,\bigg]\,\nonumber\\ 
&&{\hspace{-14cm}}-\,2av\,\bigg[\,\frac{\varepsilon^{2}}{8\pi^{2}v^{2}}\,\bigg\{\ln\,\frac{2vk_{0}}{\varepsilon}\bigg\}\,2na\,e^{-2na\rho}\,\nonumber\\ 
&&{\hspace{-12.5cm}}+\,\frac{k^{2}_{0}}{8\pi^{2}}\,e^{-2na\rho}\,\bigg\{2na\,(1\,-\,2na\rho)\,-\,\frac{\varepsilon}{v}\,\bigg[\frac{(2na\rho)^{3}}{2}\,-\,(2na\rho)^{2}\,-\,\frac{3}{4}\,2na\rho\,+\,\frac{1}{8}\bigg]\cos\theta\nonumber\\ 
&&{\hspace{-12.5cm}}-\frac{\varepsilon^{2}}{4nav^{2}}\,\bigg[\frac{(2na\rho)^{5}}{8}\,-\,\frac{13}{24}\,(2na\rho)^{4}\,-\,\frac{25}{24}\,(2na\rho)^{3}\,+\,\frac{35}{16}\,(2na\rho)^{2}\,-\,\frac{2na\rho}{8}\,+\,\frac{135}{256}\bigg]\bigg\}\nonumber \\&& 
{\hspace{-5.3cm}}+\,\frac{1}{16\pi^{3}v^{2}}\,A\,\varepsilon^{2}\,2na\,e^{-\,2na\rho}\,\bigg]\,+\,.\,.\,.\,. 
\label{e63} 
\end{eqnarray} 
and 
\begin{eqnarray} 
-\,\varepsilon\,\frac{\partial}{\partial\theta}\,(\sin\theta\,F^{\mbox{\tiny(R)}}_{1})\,-\,nav\int^{\frac{\pi}{2}}_{-\,\frac{\pi}{2}}d\phi\,\cos\phi\,(b_{\sigma}\,-\,1)\,F^{\mbox{\tiny (R)}}_{1}\,\nonumber\\ 
&&{\hspace{-8cm}}=\,\int^{\frac{\pi}{2}}_{-\,\frac{\pi}{2}}d\phi\,\cos\phi\,(b_{\sigma}\,-\,1)\bigg[\,\frac{a\varepsilon^{2}}{8\pi^{2}v}\,\ln\bigg(\frac{2vk_{0}}{\varepsilon}\bigg)\,\nonumber\\ 
&&{\hspace{-2.5cm}}-\,\frac{avk^{2}_{0}}{8\pi^{2}}\bigg\{\frac{3}{16nav}\,\varepsilon\cos\theta\,+\,\frac{135}{2048(nav)^{2}}\varepsilon^{2}\bigg\}\,\nonumber\\ 
&&{\hspace{1.5cm}}+\,\frac{a}{16\pi^{3}v}\,A\,\varepsilon^{2}\,+\,. 
\,.\,.\,.\,\bigg]\nonumber\\ 
&&{\hspace{-8cm}}=\,\frac{ak^{2}_{0}}{16\pi^{2}na}\,\varepsilon\cos\theta\,+\,.\,.\,.\,.\,, 
\label{e64} 
\end{eqnarray} 
where $b_{\sigma}$ has been defined in Eq. (\ref{e37}). The 
$A$-dependent terms in Eqs. (\ref{e63}) and (\ref{e64}) originate from 
Eqs. (\ref{e62}) and (\ref{e56}) respectively after the 
$\vec{k}$-integration is carried out. Here $A$ is the integral\footnote{We
thank the referee for pointing out an error in a previous calculation of
this integral.}   
\begin{eqnarray} 
A\,=\,\int_{0}^{2\pi}d\phi\,\cos^{2}\phi\,\ln\,\bigg[\,1\,+\,\frac{1}{4}\,\cos^{2}\phi\,\bigg]\,=\,0.53536\,.\,.\,.\,. 
\label{e65} 
\end{eqnarray} 
 
The dominant effect of the ring term on the single particle 
distribution function, i.e, $f^{\mbox{\tiny (R)}}_{1}$, can now be 
determined from Eqs. (\ref{e63}) and (\ref{e64}). It is also of some 
interest to give a crude estimate of the terms that we have 
neglected. One knows from other studies in the kinetic theory of gases 
\cite{DvB_berne_book,CC_cup_book} that excluded volume corrections to 
Boltzmann equation results are the numerically most important 
corrections, until the density of the system becomes high enough that 
the mean free path of a particle is less than the size of the particle 
itself. These excluded volume corrections are provided by the Enskog 
theory, and this theory can be applied to the Lorentz gas, as well 
\cite{vLW_physica_67}. In our case, the Enskog corrections can be 
included by replacing the density parameter $n$ by $n\,(1\,-\,\pi 
na^{2})^{-\,1}\approx n\,(1\,+\,\pi na^{2})$ in the Boltzmann 
equation. The Enskog correction affects both $\lambda_{0}$ and the 
$\varepsilon$-dependent terms in the  expressions for $\lambda_{\pm}$ 
in Eq. (\ref{e13}). Along with the Enskog correction there are other 
correction terms that affect both $\lambda_{0}$ and the 
field-dependent terms  $\lambda_{\pm}$ 
\cite{PD_unpublished,Kruis_thesis} at the same density order as the 
Enskog correction. Also, the terms that have been dropped to obtain 
Eq. (\ref{e14}) from Eq. (\ref{e10}), contribute to $\lambda_{\pm}$ 
at the same density order as the Enskog correction. However, since the 
principal objective of this paper is to investigate the  non-analytic 
contribution of the ring term  to the Lyapunov exponents, we will 
ignore the Enskog and  related corrections from our consideration. 
Thus, using Eqs. (\ref{e63}) and (\ref{e64}), one can express the full 
solutions of $f_{1}$ and $F_{1}$ as sums of  a solution in the 
Boltzmann regime, a correction due to the ring term and a correction 
due to the Enskog term, plus all of the  other terms we have 
neglected, as  
\begin{eqnarray} 
f_{1}\,=\,f^{\mbox{\tiny (B)}}_{1}\,+\,f^{\mbox{\tiny(R)}}_{1}\,+\,.\,.\,.\,. 
\label{e66} 
\end{eqnarray} 
\begin{eqnarray} 
F_{1}\,=\,F^{\mbox{\tiny (B)}}_{1}\,+\,F^{\mbox{\tiny(R)}}_{1}\,+\,.\,.\,.\,. 
\label{e67} 
\end{eqnarray} 
Consequently, for the positive Lyapunov exponent $\lambda_{+}$ we 
have,  
\begin{eqnarray} 
\lambda_{+}\,=\,\lambda^{\mbox{\tiny(B)}}_{+}\,+\,\lambda^{\mbox{\tiny (R)}}_{+}\,+\,.\,.\,.\,. 
\label{e68} 
\end{eqnarray} 
 
The solution of $F^{\mbox{\tiny (R)}}_{1}$ is quite straightforward, 
\begin{eqnarray} 
F^{\mbox{\tiny 
(R)}}_{1}\,=\,\frac{3ak^{2}_{0}}{128\pi^{2}(na)^{2}v}\,\varepsilon\cos\theta\,+\,.\,.\,.\,. 
\label{e69} 
\end{eqnarray} 
However, to solve for $f^{\mbox{\tiny (R)}}_{1}$ we find that in 
addition to the analytic field-dependent terms which can be expressed 
as a power series  in $\varepsilon$, there is a non-analytic 
field-dependent term in $f^{\mbox{\tiny (R)}}_{1}$ proportional to 
$\tilde{\varepsilon}^{2}\,\ln\,\tilde{\varepsilon}$. Thus, with  
\begin{eqnarray} 
f^{\mbox{\tiny (R)}}_{1}\,=\,f^{\mbox{\tiny (R)}}_{1,\,\mbox{\tiny analytic}}\,+\,f^{\mbox{\tiny (R)}}_{1,\,\mbox{\tiny non-analytic}}\,, 
\label{e70} 
\end{eqnarray} 
we have 
\begin{eqnarray} 
f^{\mbox{\tiny (R)}}_{1,\,\mbox{\tiny analytic}}=-\frac{ak^{2}_{0}}{4\pi^{2}}\bigg[\,2na\rho\,-\,\frac{(2na\rho)^{2}}{2}\, -\,\frac{\varepsilon\cos\theta}{2nav}\,\bigg\{\frac{(2na\rho)^{4}}{4}\,-\, (2na\rho)^{3} + \frac{(2na\rho)^{2}}{8} + \frac{2na\rho}{8}\bigg\}\,\nonumber\\ 
&&{\hspace{-12.7cm}} +\,\frac{\varepsilon^{2}}{4(nav)^{2}}\,\bigg\{-\,\frac{(2na\rho)^{6}}{32}\,+\, \frac{11}{48}\,(2na\rho)^{5}\,-\,\frac{(2na\rho)^{4}}{96}\,-\,\frac{79}{96}\,(2na\rho)^{3}\,\nonumber\\ 
&&{\hspace{-9.8cm}}+\,\frac{3}{32}\,(2na\rho)^{2}\,-\,\frac{135}{512}\,(2na\rho)\,+\,\frac{135}{512}\bigg\}\,\bigg]\,e^{-2na\rho}\nonumber\\ 
&&{\hspace{-11.8cm}}+\,\frac{a\varepsilon^{2}}{(2\pi)^{3}v^{2}}\,A\,(1\,-\,2na\rho)\,e^{-2na\rho}\,+\,.\,.\,.\,.\hspace{1.95cm} 
\mbox{for $\rho>\displaystyle{\frac{a}{2}}$}\nonumber\\ 
&&\hspace{-14.7cm}=\bigg[\frac{aA}{(2\pi)^{3}v^{2}}\,-\,\frac{135 ak^{2}_{0}}{512\, (4\pi nav)^{2}}\bigg]\bigg\{1-\sqrt{1\,-\,\big(\frac{2\rho}{a}\big)^{2}}\bigg\}\, \varepsilon^{2}\,+\,.\,.\,.\,.\hspace{1cm}\mbox{for $\rho<\displaystyle{\frac{a}{2}} 
$} 
\label{e71} 
\end{eqnarray} 
and 
\begin{eqnarray} 
f^{\mbox{\tiny (R)}}_{1,\,\mbox{\tiny non-analytic}}\,=\,\frac{a\varepsilon^{2}}{4\pi^{2}v^{2}}\,\bigg\{\ln\,\frac{2vk_{0}}{\varepsilon}\bigg\}\,(1\,-\,2na\rho)\,e^{-2na\rho}\hspace{1.8cm}\mbox{for $\rho>\displaystyle{\frac{a}{2}}$}
\nonumber\\
&&\hspace{-11.15cm}=\,\frac{a\varepsilon^{2}}{4\pi^{2}v^{2}}\,
\bigg\{\ln\,\frac{2vk_{0}}{\varepsilon}\bigg\}\bigg[\,1-\sqrt{1\,-\,
\big(\frac{2\rho}{a}\big)^{2}}\,\bigg]\hspace{1.4cm}  \mbox{for
$\rho<\displaystyle{\frac{a}{2}}$}\,. 
\label{e72} 
\end{eqnarray} 
Notice that the ring contribution to the distribution function in
Eqs. (\ref{e71}) and (\ref{e72}) satisfies the boundary conditions
that $f^{\mbox{\tiny (R)}}_{1}\rightarrow 0$ as $\rho\rightarrow0$ and
$\rho\rightarrow\infty$. Equations (\ref{e71}) and (\ref{e72}) also
satisfy continuity at $\rho=\displaystyle{\frac{a}{2}}$ at the leading
density order. The distribution functions, Eqs. (\ref{e69}),
(\ref{e71}) and (\ref{e72}), are all the ones that we need to
calculate $\lambda^{\mbox{\tiny (R)}}_{+}$. Consequently, 
\begin{eqnarray} 
\lambda^{\mbox{\tiny
(R)}}_{+}\,=\,\lambda^{\mbox{\tiny(R)}}_{+,\,\mbox{\tiny
analytic}}\,+\,\lambda^{\mbox{\tiny(R)}}_{+,\,\mbox{\tiny
non-analytic}} 
\label{e74} 
\end{eqnarray} 
and using the the definition of Lyapunov exponents in Eq. (\ref{e11}),
we have 
\begin{eqnarray} 
\lambda^{\mbox{\tiny (R)}}_{+,\,\mbox{\tiny
analytic}}\,=\,\int_{0}^{2\pi}d\theta\,\int_{\frac{a}{2}}^{\infty}d\rho\,\frac{f^{\mbox{\tiny(R)}}_{1,\,\mbox{\tiny
analytic}}}{\rho}\,\nonumber\\  &&{\hspace{-5.3cm}}=\,-\,\frac{a
k^{2}_{0} v}{4\pi}\,-\,\frac{ak^{2}_{0} l^{2} \varepsilon^{2}}{2\pi
v}\,\bigg\{\,\frac{13}{96}\,-\,\frac{135}{512}\,\big(\ln
2na^{2}\,+\,{\cal C}\,\big)\,\,\bigg\}\,\nonumber\\
&&{\hspace{-1cm}}-\,0.53536\,\frac{a\varepsilon^{2}}{(2\pi)^{2}v}\,\big(\ln\,2na^{2}\,+\,{\cal
C}\,\big)\,+\,.\,.\,.\,. 
\label{e75} 
\end{eqnarray} 
and 
\begin{eqnarray} 
\lambda^{\mbox{\tiny (R)}}_{+,\,\mbox{\tiny
non-analytic}}\,=\,\int_{0}^{2\pi}d\theta\,\int_{\frac{a}{2}}^{\infty}d\rho\,\frac{f^{\mbox{\tiny(R)}}_{1,\,\mbox{\,\tiny
non-analytic}}}{\rho}\,\nonumber\\
&&{\hspace{-6.2cm}}=\,-\,\frac{a\varepsilon^{2}}{2\pi
v}\,\bigg\{\ln\frac{2 k_{0} v}{\varepsilon}\bigg\}\,\big(\ln
2na^{2}\,+\,{\cal C}\,\big)\,, 
\label{e76} 
\end{eqnarray}  
where $l$ is the mean free path and ${\cal C}$ is Euler's constant,
${\cal C}\,=\,0.5772\,.\,.\,.\,$.

\subsection{Calculation of $\lambda^{\mbox{\tiny (R)}}_{-}$} 
 
To calculate the corresponding effect of the ring term on
$\lambda_{-}$, we make use of the relation Eq. (\ref{e12}). It is easy
to calculate the effect of the ring term on
$\big<\alpha\big>_{\mbox{\tiny NESS}}$ using $F^{\mbox{\tiny
(R)}}_{1}$ already determined in the previous section. Thus, using  
\begin{eqnarray} 
\lambda_{+}\,+\,\lambda_{-}\,=\,-\,\big<\alpha\big>_{\mbox{\tiny
NESS}}\,, 
\label{e88} 
\end{eqnarray} 
and a complete analogy to Eqs. (\ref{e66}-\ref{e68}), we can calculate
three terms of $\big<\alpha\big>_{\mbox{\tiny NESS}}$: 
\begin{eqnarray} 
\big<\alpha\big>_{\mbox{\tiny NESS}}\,=\,\big<\alpha\big>^{\mbox{\tiny
(B)}}_{\mbox{\tiny NESS}}\,+\,\big<\alpha\big>^{\mbox{\tiny
(R)}}_{\mbox{\tiny NESS}}\,+\,.\,.\,.\,.\,, 
\label{e89} 
\end{eqnarray} 
with 
\begin{eqnarray} 
\big<\alpha\big>^{\mbox{\tiny (R)}}_{\mbox{\tiny
NESS}}\,=\,\frac{3ak^{2}_{0} l^{2} \varepsilon^{2}}{32\pi
v}\,+\,\,.\,.\,.\,. 
\label{e90} 
\end{eqnarray} 
 
Following Eqs. (\ref{e66}-\ref{e68}), we now express $\lambda_{-}$ as
$\lambda_{-}\,=\,\lambda^{\mbox{\tiny(B)}}_{-}\,+\,\lambda^{\mbox{\tiny
(R)}}_{-}\,+\,.\,.\,.\,.\,$,  satisfying $\lambda^{\mbox{\tiny
(R)}}_{+}\,+\,\lambda^{\mbox{\tiny(R)}}_{-}\,=\,-\,\big<\alpha\big>^{\mbox{\tiny
(R)}}_{\mbox{\tiny NESS}}$.   This leads us to 
\begin{eqnarray} 
\lambda^{\mbox{\tiny (R)}}_{-,\,\mbox{\tiny
analytic}}\,=\,\frac{ak^{2}_{0} v}{4\pi}\,-\,\frac{a k^{2}_{0} l^{2}
\varepsilon^{2}}{2\pi
v}\,\bigg\{\,\frac{5}{96}\,+\,\frac{135}{512}\,(\,\ln2na^{2}\,+\,{\cal
C}\,\big)\,\bigg\}\,\nonumber\\&&{\hspace{-6cm}}+\,\,0.53536\,\frac{a\varepsilon^{2}}{(2\pi)^{2}v}\,(\,\ln\,2na^{2}\,+\,{\cal
C}\,\big)\,+\,.\,.\,.\,.\,, 
\label{e92} 
\end{eqnarray} 
and 
\begin{eqnarray} 
\lambda^{\mbox{\tiny (R)}}_{-,\,\mbox{\tiny
non-analytic}}\,=\,\frac{a\varepsilon^{2}}{2\pi v}\bigg\{\ln
\frac{2k_{0} v}{\varepsilon}\bigg\}\,(\,\ln 2na^{2}\,+\,{\cal
C}\,\big). 
\label{e93} 
\end{eqnarray} 
where $l$ is the mean free path and ${\cal C}$ is Euler's constant,
${\cal C}\,=\,0.5772\,.\,.\,.\,$. 
 
\section{The field-dependent collision frequency and its \\effects on  
the Lyapunov exponents} 
 
As stated before, our second main purpose was the derivation of the
leading non-analyticity in the field dependence of the Lyapunov
exponents. In analogy with the transport coefficients, we expected
these  non-analyticities to result from the long time behavior of the
ring  terms, which we found confirmed in the preceding section. Some
further thought reveals we can estimate the non-analytic field
dependence in a simple  way. 
 
In the presence of a thermostatted field there are two types of
contributions to the positive Lyapunov exponent of the two-dimensional
Lorentz gas:\\ 1)  contributions from the bending of the trajectories
by the fields and\\ 2) contributions from the divergence of trajectory
pairs at  collisions. 
 
The first type of contributions are of order $\tilde{\varepsilon}^2$
in the Boltzmann approximation. We expect that the coefficient of this
term will pick up higher density corrections and there will be
additional terms of higher orders in $\tilde{\varepsilon}$. But we
have not found any indications for corrections of lower order than
$\tilde{\varepsilon}^2$ resulting from the field-bending
contributions. 
 
The collisional contributions can be  generally expressed as an
average of the form
$\nu\,\langle\,\ln\displaystyle{\frac{|\delta\vec{v'}|}{|\delta\vec{v}|}}\,\rangle_{{\mbox{\scriptsize
c}}}$, with $\delta \vec{v'}$ and $\delta \vec{v}$ the velocity
differences between the  adjacent trajectories just after and just
before a collision, respectively, $\nu$ the average collision
frequency, and the angular brackets,
$\langle\,\rangle_{{\mbox{\scriptsize c}}}$, indicating an average
over  collisions. At low densities even correlated collisions happen
at large distances, i.e. in the order of a mean free path length apart
from each other. Therefore their distribution of collision angles and 
hence their contribution to the average
$\langle\,\rangle_{{\mbox{\scriptsize c}}}$, to the leading order in
density remains the same as for  uncorrelated collisions. We should
then expect that at low densities  the main effect of the correlated
collisions on the Lyapunov exponents  should be due to a change of the
collision frequency $\nu$ as a result  of correlated collisions taking
place in the presence of the field. If  the  latter changes from
$\nu_0$ to $\nu_0+\delta\nu$, then  Eq. (\ref{e131}) predicts a change
of the positive Lyapunov exponent of  magnitude 
\begin{eqnarray} 
\delta\lambda_{+}\,=\,-\,\delta\nu\,\bigg\{\ln
\frac{a\nu_{0}}{v}\,+\cal{C}\bigg\}. 
\label{deltalambda} 
\end{eqnarray} 
To obtain this result we have used the fact that the equilibrium, low
density Lyapunov exponent, Eq. \ref{e131}) can be written in the form 
\begin{eqnarray} 
\lambda_0\,=\,\nu_{0}\,\bigg\{1\,-\,{\cal
C}\,-\,\ln\frac{a\nu_{0}}{v}\bigg\}, 
\label{e1311} 
\end{eqnarray} 
where $\nu_{0}=2nav$. In order to understand why and how the
thermostatted field changes  the collision frequency we first recall
that in  equilibrium the collision  frequency can be obtained simply
by using the uniformity of the equilibrium  distribution for the point
particle in available phase space, with the result that
$\nu=\displaystyle{\frac{2nav}{1-\pi n a^2}}$.  One just has to
consider the probability that the light particle   during an
infinitesimal time $d\,t$ will hit one of the scatterers. On the other
hand, at a time $t$ after a given initial time, the probability for a
collision may be considered to be a sum of three  contributions: the
collision frequency obtained by assuming that all  collisions are
uncorrelated and independent of each other, {\em plus}  the
probability for a recollision with a scatterer with which it has
collided before, {\em minus} the  reduction of the collision
probability due to  any collected knowledge  of where no scatterers
are present.  In equilibrium the last two contributions have to
cancel, as we demonstrate in the Appendix. In the presence of a field,
however, this cancellation does not occur. This can easily be
understood in a qualitative way following the argument that the
cancellation in equilibrium occurs because the probability for return
to the boundary of a scatterer is exactly the same as that for return
to the boundary of a region where a scatterer could be, but in fact is
not present (a virtual scatterer). In the presence of a field, the
average velocity of the point particle before collision with a real
scatterer will be in the direction of the field, and after the
collision the average velocity  will be anti-parallel to the
field. The  field will then tend to turn the particle around and have
it move back  in the direction of the scatterer. This effect enhances
the  probability of a recollision in comparison to that for an
isotropic  distribution around the scatterer. In a ``virtual
collision", in which  the velocity does not change, the particle, on
average, ends up  downstream (i.e. in the direction of the applied
field) from the  virtual scatterer and its recollision  probability is
decreased  compared to that for an isotropic distribution.  
 
In the Appendix, a quantitative calculation is given based on the
following two assumptions:\\ 1) After the real or virtual collision
the spatial distribution of the point particle becomes centered around
a point at a distance of a diffusion length from the scatterer
and\\ 2) for long times this distribution can be found by solving  the
diffusion equation. The resulting expression for $\delta\nu$ is 
\begin{eqnarray} 
\delta\nu\,=\,\frac{a\varepsilon^2}{2\pi
v}\ln\frac{\nu_0}{\varepsilon}\,. 
\label{delnu} 
\end{eqnarray} 
 
A more formal, but equivalent, way to obtain this result is by
extending the method  described by Latz, van Beijeren and Dorfman
\cite{LvBD_preprint} for the low density distribution of time of free
flights of the moving particle to include the contribution from ring
events, so as to apply to a system in a thermostatted electric
field. The main idea is to solve a kinetic equation for
$f(\vec{r},\vec{v},t,\tau)$, the distribution of particles at a phase
point $(\vec{r},\vec{v})$ at time $t$ such that their last collision
took place at a time $\tau$ earlier, i.e., at time $t-\tau$. It is
then easy to argue that the distribution of free flight times is
simply the derivative of this (``last collision'') distribution with
respect to $t-\tau$.  We can then obtain a NESS average of the time 
of free flight and thereby calculate the field dependent collision frequency
$\nu(\varepsilon)=\nu_{0}+\delta\nu$. Since we want to show that the
origin of the non-analytic field dependence of both $\lambda_{+}$ and
$\lambda_{-}$  is rooted in the non-analytic field dependence of
collision frequency $\delta\nu$, let us keep only the non-analytic
field-dependent term as the leading term of the expansion of
$\delta\nu$ in the  density of scatterers and in the electric field
strength and write 
\begin{eqnarray} 
\delta\nu\,=\,\beta\,{\varepsilon^{2}}\ln\bigg\{\frac{2k_{0}v}{\varepsilon}\bigg\}\,+\,\cdots\,, 
\label{nu1} 
\end{eqnarray} 
where the quantity $\beta$ has to be determined from the NESS average
of $\tau$, using the effect of the ring term on the NESS distribution
function $f(\vec{r},\vec{v},\tau)$ with $k_{0}$ of the order of
$\displaystyle{\frac{1}{\nu_{0}v}}\,$. To obtain this distribution
function, we follow exactly the same procedure as outlined in Sections
4 and 5, but this time, with the variable $\tau$ instead of
$\rho$. Notice that, this time, even though the equations for
corresponding $f_{1}$ and $g_{2}$'s are different, due to the
difference in the dynamical equations for $\dot{\rho}$ and
$\dot{\tau}$ during free flights and at collisions, the equations
involving $F_{1}$ and $G_{2}$'s remain the same. The source of the
non-analytic field-dependent term will surface again exactly from the
$O(k^{-2})$ term in Eq. (\ref{e54}). As far as this non-analytic
field-dependent term is concerned, at the lowest order of density, the
variables $\rho$ and $\tau$ are identical up to a multiplicative
factor $v$. Both grow linearly with time in between collisions and are
set back to (for $\rho$, almost) zero at each collision with a
scatterer. One then recovers the corresponding non-analytic part of
the NESS distribution function \cite{Panja_thesis}, analogous to
Eq. (\ref{e72}),  
\begin{eqnarray} 
f^{\mbox{\tiny (R)}}_{1,\,\mbox{\tiny 
non-analytic}}(\vec{v},\tau)\,=\,\frac{a\varepsilon^{2}}{4\pi^{2}v}\,\bigg\{\ln\,\frac{2vk_{0}}{\varepsilon}\bigg\}\,(1\,-\,2nav\tau)\,e^{-2nav\tau}\hspace{1.8cm}\mbox{for\, $\tau>0$}\,, 
\label{nu2} 
\end{eqnarray} 
from which $\beta$ can be obtained to be 
\begin{eqnarray} 
\beta\,=\,\frac{a}{2\pi v}\,, 
\label{nu3} 
\end{eqnarray} 
after which, one easily recovers the result of Eq. (\ref{delnu}).   
 
\section{Discussion} 
 
While much of this paper is quite technical, there are two main points
that we would like to emphasize: (1) We have developed a method which
allows an extension of the calculation of the Lyapunov exponents for a
two-dimensional Lorentz gas to higher densities than is possible by
means of the ELBE. (2) The logarithmic terms obtained here, while
small, are indicators of similar logarithmic terms which are certain
to appear when these calculations are extended to general
two-dimensional gases, where all of the particles move.  
 
The first point allows one to contemplate a general kinetic theory for
the calculation of of sums, at least, of all positive, or of all
negative Lyapunov exponents. Such an approach was also indicated by
Dorfman, Latz, and van Beijeren \cite{DLvB_chaos_98}, for the
KS-entropy of a dilute gas in equilibrium, but the theory there has
not yet been developed beyond the Boltzmann equation.  The relevance
of the second point can be seen if one realizes that the linear
Navier-Stokes transport coefficients of a two-dimensional gas diverge
because of long time tail effects, of the type discussed here
\cite{DvB_berne_book}. In the general gas case therefore the
logarithmic terms in the positive and negative Lyapunov exponents will
not cancel as they do here, because the transport coefficients
themselves should diverge as $\ln\tilde{\varepsilon}$ as $\varepsilon$
approaches zero. Thus the logarithmic terms obtained here should be
seen as precursors of the more important logarithmic terms that will
appear in the theory of two-dimensional gases. 
 
It is worth noting that the
$\tilde{\varepsilon}^{2}\ln{\tilde{\varepsilon}}$ term results from a
long range correlation in time between the moving particle and the
scatterers that is present in both the pair correlation functions,
$G_2$, and $g_2$,  either of which  is proportional to the square of
the electric field strength and the inverse square of the wave number,
at small wave numbers and fields. This dependence is not present in
the  Lorentz gas in equilibrium, of course, but similar collision
frequency arguments to those given here suggest that non-analytic
terms may be present in the ring contributions to the positive
Lyapunov exponent for trajectories on the fractal repeller for an open
Lorentz gas. In this case the inverse system size, $L^{-1}$, plays the
role of $\displaystyle{\frac{\varepsilon}{2v}}$, the lower limit of
$k$ for the integration over $\vec{k}$ and  one would expect to find
terms of order $L^{-2}\ln L$ in the ring term for this case. This
point is currently under investigation. 
 
Finally we mention that neither the non-analytic terms found here, nor
the excluded volume corrections included in the Enskog terms are able
to account for the field dependence of the Lyapunov exponents as
observed in the computer simulations by Dellago and Posch
\cite{vBDCPD_prl_96}. This is not unexpected since we have not been
systematic in computing the density dependence of the coefficient of
$\varepsilon^{2}$, nor have we considered higher order terms in
$\varepsilon$ beyond order $\varepsilon^{2}\ln\varepsilon$. All of the
neglected terms are likely to be numerically more important than the
ones we have kept. There is also no indication in the simulation data
for the Lyapunov exponents of a clear presence of the interesting
logarithmic term in the applied field. Such logarithmic terms are
typically difficult to detect in simulation data, without a careful
hunt for them\cite{vLW_physica_67}.  However, it may be easier to
check, by means of computer simulation, the  existence of the
$\tilde{\varepsilon}^{2}\ln{\tilde{\varepsilon}}$ term in the
collision frequency than in the Lyapunov exponents. In any case, we
would like to emphasize that computer simulation studies of
thermostatted systems provide very useful ways to check a number of
phenomena predicted by the kinetic theory of moderately dense gases.   
 
ACKNOWLEDGEMENTS: This paper is dedicated to George Stell on the
occasion of his 60-th birthday. We would like to thank Kosei Ide,
Zolt\'{a}n Kov\'{a}cs, Herman Kruis, Arnulf Latz, Luis Nasser, and
Ramses van Zon for many valuable conversations and suggestions. J.\
R.\ D.\ wishes to thank the National Science  Foundation for support
under Grant PHY-96-00428. H.\ v.\ B.\  acknowledges support by FOM,
SMC and by the NWO Priority Program Non-Linear Systems, which are
financially supported by the "Nederlandse Organisatie voor
Wetenschappelijk Onderzoek (NWO)".

\appendix 
 
\section*{Appendix\\Derivation of the field-dependent collision  
frequency} 
 
\setcounter{equation}{0} 
 
\renewcommand{\theequation}{A\arabic{equation}} 
 
To derive the field dependence of the collision frequency we first
approximate the probability of a recollision at time $t$ as 
\begin{eqnarray} 
P^{\mbox{\scriptsize
rec}}(t)\,=\,\frac{\nu}{2}\,\int_0^{2\pi}d\,\theta\int_0^{\infty}d\tau\,\int_{\vec{v}\cdot\hat{\sigma}>0}d\hat{\sigma}\,|\vec{v}\cdot
\hat{\sigma}|\,R(\tau,\theta,\sigma)\,b_{\hat{\sigma}}\,F^{\mbox{\tiny(B)}}(\theta)\,. 
\label{Prec} 
\end{eqnarray} 
Here $F^{\mbox{\tiny (B)}}(\theta)$ describes the Boltzmann
distribution for the velocity in the NESS. The function
$R(\tau,\theta,\sigma)$ describes the probability density for return
to the circumference of a given scatterer in a time $\tau$ just after
colliding with this scatterer with scattering vector $\hat{\sigma}$
and post-collisional velocity described by $\theta$ (see  Fig. 4). We
have ignored a possible dependence of the collision frequency $\nu$ on
$\hat{v}$, which would only play a role at higher orders in the
density. 
 
\vspace{5mm} 
\hspace{0.8in}{\includegraphics[width=4.267in]{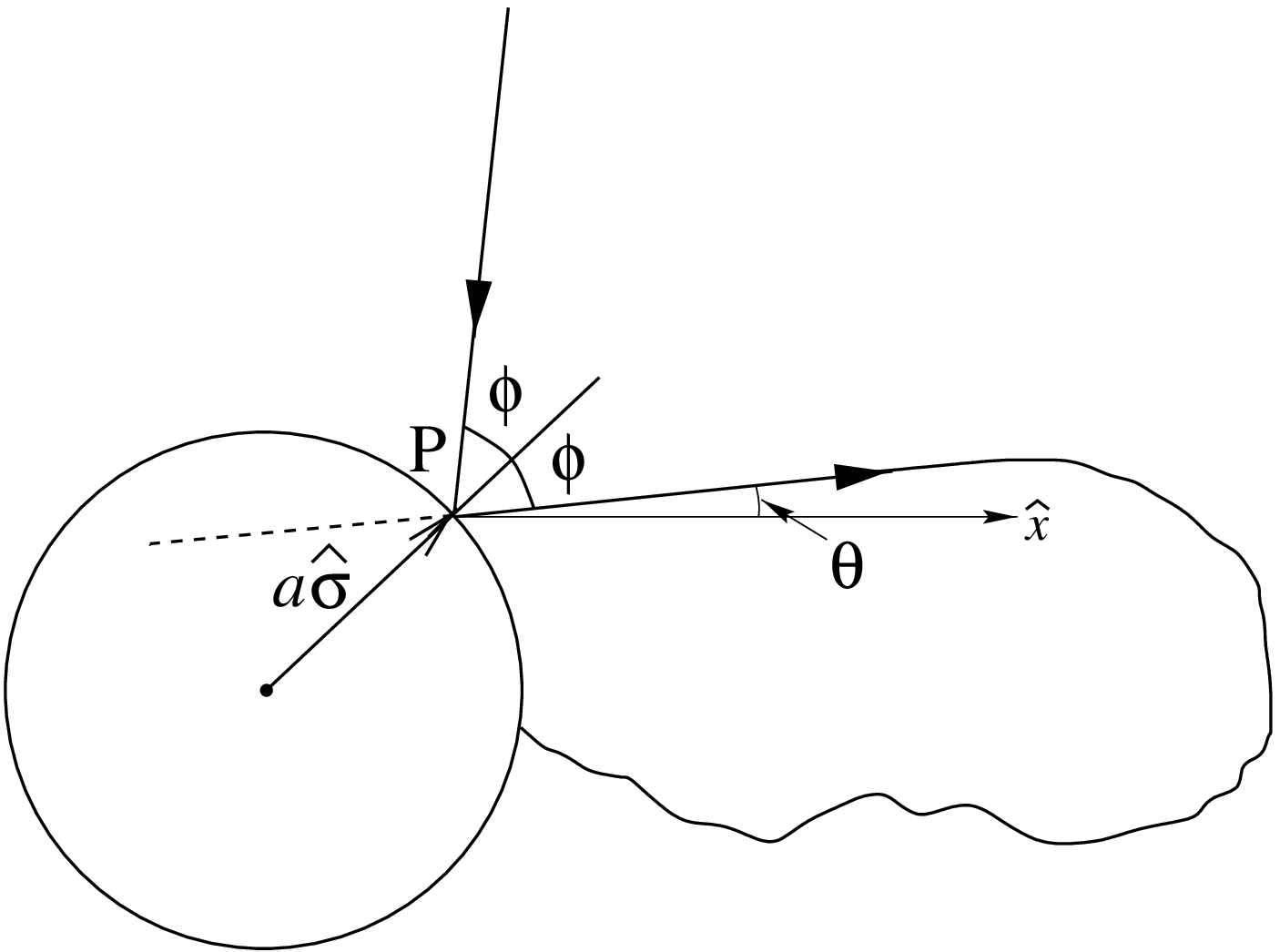}}
\vspace{-3mm} 
\begin{center} 
Fig. 4 : A recollision taking place after a real (solid line)   or
corresponding virtual (dashed line) collision, followed by a
post-collisional excursion maintaining on average the direction of
velocity  over a persistence length $l_p$. 
\end{center} 

Similarly the reduction of the collision frequency at time $t$ due to
virtual recollisions can be estimated as 
\begin{equation} 
P^{\mbox{\scriptsize nc}}(t)\,=\,-\,\frac{\nu}{2}\int_0^{2\pi}
d\theta\int_0^{\infty}d\tau\int_{\vec{v}\cdot\hat{\sigma}>0}d\hat{\sigma}\,|\vec{v}\cdot
\hat{\sigma}|\,R(\tau,\theta,\sigma)\,F^{\mbox{\tiny (B)}}(\theta)\,. 
\label{Prec1} 
\end{equation} 
 
In equilibrium $F^{\mbox{\tiny (B)}}(\theta)$ is independent of
$\theta$, so one sees immediately that both terms cancel, as they
should. In the presence of a thermostatted field we need the explicit
form of $F^{\mbox{\tiny (B)}}_{1}(\theta)$ up to the first field-dependent
order, given in Eq. (\ref{e50}) as 
\begin{equation} 
F^{\mbox{\tiny(B)}}_{1}(\theta)\,=\,\frac{1}{2\pi}\left[\,1\,+\,\frac{3\varepsilon}{8nav}\,\cos\theta\,\right]. 
\label{A2} 
\end{equation} 
The function $R(\tau,\theta,\sigma)$ for large enough $\tau$ may be
approximated by the product of $2av$ (velocity times cross section)
and the probability density for finding the point particle at the
position of the scatterer. For weak fields the latter may be
approximated by the solution of a diffusion equation with a drift
velocity $u\hat{x}$ in  the $+x$-direction and an initial density
localized at the position  $l_p \hat{\theta}$ with respect to the
center of the scatterer. Here  $l_p$  is the persistence length, that
is,  the average distance traveled by a point particle  in an
equilibrium system in the direction of  its initial velocity and
$\hat{\theta}$ is the unit vector in the  direction of the velocity
right after the initial collision at  $t-\tau$. The persistence length
may be expressed as  $
l_{p}=\displaystyle{\int_0^{\infty}dt\,\langle\hat{v}\cdot
\vec{v}(t)\rangle}$. Multiplying this by the constant speed $v$ one
finds with the aid of the Green-Kubo expression for the diffusion that
$l_p=\displaystyle{\frac{2D}{v}}$ in two dimensions. This assumption
for the long time distribution may be understood by imagining that the
first few free flights after the initial collision of the particle
move it over a distance in the order of a mean free path in the
direction of its initial postcollisional velocity before it starts to
diffuse by virtue of further collisions with scatterers. Thus for
large $\tau$ the distribution of the light particle will be centered
around the point $l_p\hat{\theta}$ with respect to the center of the
scatterer, and the final point, on the surface of the scatterer, may
be approximated to be at the center of the scatterer as well, because
of low density. These arguments lead to the explicit form for the
recollision probability given by 
\begin{equation} 
R(\tau,\theta,\sigma)\,=\,2av\,\frac{e^{-\frac{[\,l_p\hat{\theta}\,+\,u\tau\hat{x}\,]^{2}}{4D\tau}}}{4\pi
D\tau}\,. 
\label{A3} 
\end{equation} 
Finally we need the explicit form $u=\displaystyle{\frac{3\varepsilon
v}{8\nu_0}}$ for the drift velocity to leading order in the density,
and the identity that 
\begin{equation} 
\frac{1}{2}\,\int_{\vec{v}\cdot\hat{\sigma}>0}d\hat{\sigma}\,|\vec{v}\cdot\hat{\sigma}|\,b_{\hat{\sigma}}\,\cos\theta\,=\,-\,\frac{v}{3}\,\cos\theta\,. 
\end{equation} 
Then, after expanding 
\begin{equation} 
e^{-\,\frac{2l_p
u\tau\hat{\theta}\cdot\hat{x}}{4D\tau}}\,=\,1\,-\,\frac{l_p
u}{2D}\cos\theta\,+\,.\,.\,.\,, 
\end{equation} 
we can now do all the calculations needed to obtain the leading
non-analytic term in the field expansion of the collision frequency.
We find that 
\begin{eqnarray} 
\delta\nu\,=\,av\nu\int_{0}^{2\pi}d\theta\int_{0}^{\infty}d\tau\,\frac{e^{-\frac{[\,l_{p}^{2}\,+\,(u\tau)^{2}]}{4D\tau}}}{4\pi
D\tau}\,\bigg[\,1\,-\,\frac{l_{p}u}{2D}\,\cos\theta\,\bigg]\int_{\vec{v}\cdot{\hat{\sigma}}>0}d\hat{\sigma}\,|\vec{v}\cdot\hat{\sigma}|\,
(b_{\hat{\sigma}}-1)\,\frac{3\varepsilon\cos\theta}{16\pi nav}\,. 
\label{A4} 
\end{eqnarray} 
After performing the integrations, we recover
Eq. (\ref{delnu}). Notice that the   logarithm of
$\tilde{\varepsilon}$ results from the cut-off on the $\tau$
integration provided by the drift term in the exponential.

\end{document}